%% file: root.tex
\documentclass[twocolumn]{autart_arXiv}

\usepackage{graphicx}          

\usepackage[utf8]{inputenc}
\usepackage{svg}
\usepackage{booktabs}
\usepackage[english]{babel}
\usepackage{amsmath}
\usepackage{amssymb}
\usepackage{mathtools}
\usepackage{physics}
\usepackage{dirtytalk}
\usepackage{booktabs}
\usepackage{comment}
\usepackage{caption}
\usepackage{subcaption}
\usepackage{enumitem}
\usepackage{url}

\definecolor{matlab1}{rgb}{0 0.4470 0.7410}
\definecolor{matlab2}{rgb}{0.8500 0.3250 0.0980}
\definecolor{matlab3}{rgb}{0.9290 0.6940 0.1250}
\definecolor{matlab4}{rgb}{0.4940 0.1840 0.5560}
\definecolor{matlab5}{rgb}{0.4660 0.6740 0.1880}
\definecolor{matlab6}{rgb}{0.3010 0.7450 0.9330}
\definecolor{matlab7}{rgb}{0.6350 0.0780 0.1840}

\usepackage[hidelinks]{hyperref}

\usepackage[capitalise, nameinlink, noabbrev]{cleveref}
\crefname{equation}{}{}
\Crefname{equation}{Equation}{Equations}
\crefname{assumption}{Assumption}{Assumptions}
\Crefname{assumption}{Assumption}{Assumptions}
\crefname{thm}{Theorem}{Theorems}
\Crefname{thm}{Theorem}{Theorems}
\crefname{lemma}{Lemma}{Lemmas}
\Crefname{lemma}{Lemma}{Lemmas}
\crefname{remark}{Remark}{Remarks}
\Crefname{remark}{Remark}{Remarks}

\newcommand{\sign}[1]{\operatorname{sign}\left(#1\right)}
\newcommand{\ltwonorm}[1]{\norm{#1}_{L_2}}
\newcommand{\lonenorm}[1]{\norm{#1}_{L_1}}
\newcommand{\alignbreak}{\nonumber\\&\quad~}
\renewcommand{\cp}{c_\text{p}}
\newcommand{\Ta}{T_\text{a}}
\newcommand{\thetab}{\bar{\theta}}

\newcommand{\Acs}{A_{\text{cs}}}
\newcommand{\Qtem}[1]{\dot{Q}_{u,#1}}
\newcommand{\Qloss}[1]{\dot{Q}_{\text{loss},#1}}
\newcommand{\Qpsi}[1]{\dot{Q}_{\psi,#1}}
\newcommand{\qpsi}[1]{\dot{q}_{\psi,#1}}
\newcommand{\hal}{h_{\text{al}}}
\newcommand{\bal}{b_{\text{al}}}
\newcommand{\Lal}{L_{\text{al}}}
\newcommand{\ltem}{l_\text{TEM}}
\newcommand{\Atemloss}{A_{\text{a}}}
\newcommand{\propgain}{k}
\newcommand{\Te}{T_\text{e}}
\newcommand{\Ten}{T_{\text{e},0}}
\newcommand{\TeL}{T_{\text{e},L}}
\newcommand{\ze}{\vartheta_\text{e}}
\newcommand{\une}{u_{\text{e},0}}
\newcommand{\uoe}{u_{\text{e},1}}
\newcommand{\errvar}{\bar{z}}
\newcommand{\bb}{\bar{b}}
\newcommand{\ltfac}{\sqrt{\frac{|\lambda|}{\thetab}}}
\newcommand{\Phid}[1]{\Phi_{\mathrm{d},#1}}
\newcommand{\bm}[1]{b_{\mathrm{m},#1}}
\newcommand{\bM}[1]{b_{\mathrm{M},#1}}
\newcommand{\am}[1]{a_{\mathrm{m},#1}}
\newcommand{\aM}[1]{a_{\mathrm{M},#1}}
\newcommand{\thetam}{\theta_{\mathrm{m}}}
\newcommand{\ka}[1]{\bar{k}_{#1}}
\newcommand{\lM}{\lambda_\mathrm{M}}
\newcommand{\lm}{\lambda_\mathrm{m}}

\newcommand{\Rth}[1]{R_{\theta,#1}}

\edef\endfrontmatter{\unexpanded\expandafter{\endfrontmatter}\noexpand\endNoHyper }

\newtheorem{assumption}{Assumption}
\newtheorem{remark}{Remark}
\newtheorem{lemma}{Lemma}

\allowdisplaybreaks 

\begin{document}

\begin{frontmatter}

\title{Robust Adaptive Boundary Control of a Thermal Process with  Thermoelectric Actuators: Theory and Experimental Validation\thanksref{footnoteinfo}}

\thanks[footnoteinfo]{Corresponding author: Paul Mayr.}

\author[Graz]{Paul Mayr}\ead{paul.mayr@tugraz.at},    \author[Cagliari]{Alessandro Pisano}\ead{apisano@unica.it},               \author[Graz]{Stefan Koch}\ead{stefan.koch@tugraz.at},  \author[Graz]{Markus Reichhartinger}\ead{markus.reichhartinger@tugraz.at}

\address[Graz]{Graz University of Technology, Institute of Automation and Control, Graz, Austria.}  \address[Cagliari]{University of Cagliari, Department of Electrical and Electronic Engineering, Cagliari, Italy.}


\begin{abstract}                          A sliding-mode--based adaptive boundary control law is proposed for a class of uncertain thermal reaction-diffusion processes subject to matched disturbances. The disturbances are assumed to be bounded, but the corresponding bounds are unknown, thus motivating the use of adaptive control strategies. A boundary control law comprising a proportional and discontinuous term is proposed, wherein the magnitude of the discontinuous relay term is adjusted via a gradient-based adaptation algorithm. Depending on how the adaptation algorithm is parameterized, the adaptive gain can be either a nondecreasing function of time (monodirectional adaptation) or it can both increase and decrease (bidirectional adaptation). The convergence and stability properties of these two solutions are investigated by Lyapunov analyses, and two distinct stability results are derived, namely, asymptotic stability for the monodirectional adaptation and globally uniformly ultimately bounded solutions for the bidirectional adaptation. The proposed algorithms are then specified to address the control problem of stabilizing a desired temperature profile in a metal beam equipped with thermoelectric boundary actuators. Experiments are conducted to investigate the real-world performance of the proposed sliding-mode-based adaptive control, with a particular focus on comparing the monodirectional and bidirectional adaptation laws.\end{abstract}

\end{frontmatter}

\section{Introduction}
Partial differential equations (PDEs) are central to modelling key industrial and physical processes. Diffusion-type PDEs are particularly used to characterize industrial processes and components such as tubular reactors \cite{Boskovic2002}, lithium-ion batteries \cite{Forman2011} and distillation processes \cite{Eleiwi2017}, to name a few. In many of these systems, actuation is naturally confined to the boundaries of the spatial domain \cite{Burns2016,Ng2013}. Furthermore, these systems often operate under severe uncertainty caused, e.g., by unpredictable spatial variability in material properties,   external disturbances, and actuator imperfections. This necessitates the use of \emph{robust} control schemes which guarantee the desired behaviour despite the presence of these effects. If disturbances are matched, i.e., they appear in the same channel as the manipulable control input, a popular robust control technique is sliding-mode control. Ongoing research is concerned with the application of sliding-mode--based controllers on several classes of systems governed by PDEs \cite{Pisano2017,Wang2019,Koch2022,Zhou2025,Balogoun2025,Zhang2025}. Typically, sliding-mode--based controllers are designed to compensate for disturbances which are bounded in magnitude and for which the corresponding bound is known in advance. If the bound is unknown, adaptive strategies can be employed to adjust the controller gains online. See, e.g., \cite{Plestan2010} for an outline of this class of adaptive control algorithms in the finite-dimensional setting. These algorithms can generally be divided into two groups: monodirectional and bidirectional adaptation. In the former case, the adaptive gain increases over time until it becomes large enough to compensate for the disturbance. The gain is not allowed to decrease again, even when the amplitude of the disturbance possibly decreases. An algorithm of this form is applied to the diffusion PDE in \cite{Mayr2024}, where it is shown that the adaptive gain remains bounded and the closed-loop system is asymptotically stabilized in the $L_2$-sense. \\
In contrast to that, \cite{Roy2020} proposes a bidirectional algorithm, that is, the adaptive gain can increase \emph{and} decrease. This is beneficial for implementation purposes and is less susceptible to over-estimation, which in turn helps alleviating the chattering phenomenon. However, as it is typical with algorithms of this type, only practical convergence towards a vicinity of zero can be guaranteed. This is also the case in, e.g. \cite{Han2024}, where a flexible string and an Euler-Bernoulli beam are stabilized via a controller which employs a bidirectional adaptation algorithm. It should be noted that due to practical implementation aspects such as measurement noise or discretization chattering, monodirectional adaptation algorithms undergo, in their ideal formulation, unbounded drift of the adaptive gains. This problem is commonly solved by modifying the adaptation law through the use of a dead-zone in the vicinity of zero. Hence, in real-world applications, the asymptotic stability result of monodirectional adaptation algorithms is always lost. \\
In this article, an adaptive sliding-mode--based boundary controller is proposed to stabilize the origin of a perturbed diffusion process with unknown spatially-varying diffusion and reaction coefficients. The reaction coefficient is allowed to be positive, which has a destabilizing effect and therefore increases the complexity of the stabilization problem. The system is controlled via actuators entering the plant dynamics via Robin boundary conditions with unknown parameters, along with matched disturbances with unknown upper bounds. The control law is comprised of a proportional and discontinuous part, wherein the magnitude of the discontinuous relay term is adjusted via a gradient-based adaptation algorithm. Based on the parametrization of the algorithm, it is either of monodirectional or bidirectional type. This leads to two distinct stability results which are proven via dedicated Lyapunov-based analyses. \\
In addition to the theoretical achievements, the real-world applicability of the proposed control scheme is demonstrated experimentally. This is done by addressing the problem of stabilizing the temperature profile of a perturbed metallic beam thermally actuated by thermoelectric modules (TEMs) realized by Peltier elements (see, e.g., \cite{Lineykin2007}) located at the two ends of the beam. TEMs are solid-state devices capable of both heating and cooling depending on the direction of applied electrical current. The considered application involves several uncertainty factors that motivate the use of the robust adaptive control algorithm. Additionally, it necessitates extending the class of systems previously dealt with in \cite{Mayr2024} by including a reaction term to model heat losses to the environment (or, e.g., exothermic reactions) and Robin boundary conditions, which will eventually emerge from the modeling of the metallic beam equipped with the TEM actuators. \\
Both the previously mentioned monodirectional and bidirectional adaptation algorithms are implemented and tested on the given experimental setup, showing that the bidirectional adaptation effectively reduces chattering.  

\subsection{Contribution and paper structure}
Expanding the class of processes compared to that considered in \cite{Mayr2024}, and building a comprehensive analysis of the closed-loop system stability properties with both the monodirectional and bidirectional adaptation in the considered PDE setting, constitute the main theoretical contributions of this work. In addition, the application of the algorithm to solve a challenging thermal control problem, and its experimental validation on a real-world setup, are provided.
The paper is structured as follows. After introducing the adopted notation, the considered class of systems is presented in \cref{sec:problem_formulation}, along with assumptions on the system's parameters and disturbance signals. It is followed by stating the proposed control and adaptation laws in \cref{sec:controller}, together with the main theoretical result consisting of a stability proof for both the mono- and bidirectional adaptation case. In \cref{sec:application}, the laboratory setup is presented, its mathematical model is derived, and closed-loop experiments are discussed to showcase the real-world applicability and to compare both adaptive algorithms. The results are summarized in \cref{sec:conclusions}.

\subsection{Notation}
The notation used throughout the paper is fairly standard. $H^{\ell}(0,1)$, with  $\ell=0,1,2,\ldots$, denotes the Sobolev space of scalar functions $f(\zeta)$ on $[0,1]$ with square
integrable derivatives $f^{(i)}(\zeta)$ up to the order $\ell$ and the $H^\ell$-norm
\begin{equation}\label{normdef}
	\|f(\cdot)\|_{H^\ell} = \sqrt{\int_0^1 \sum_{i=0}^\ell [f^{(i)} (\zeta)]^2 ~\dd \zeta}~. \nonumber
\end{equation}
We shall also utilize the standard notations $H^{0}(0,1)=L_2(0,1)$ and $\|f(\cdot)\|_{H^0} = \ltwonorm{f(\cdot)}$. Partial derivatives are indicated by indices, e.g, $z_t(x,t) = \pdv{z(x,t)}{t}$ or $z_{xx}(x,t) = \pdv[2]{z(x,t)}{x}$ 
whereas total time derivatives are written as $\dot{V}(h(t)) = \dv{V(h(t))}{t}$.
Derivatives of functions depending on one spatial variable only are indicated by $\varphi'(x) = \dv{\varphi(x)}{x}$.

\section{Problem formulation} \label{sec:problem_formulation}
Consider the space- and time-varying scalar field $z(x, t)$, evolving in the space $L_2(0,1)$, with the spatial variable $x \in [0,1]$  and time variable $t \geq 0$. Let it be governed by the perturbed initial-boundary-value problem given by the PDE
\begin{subequations} \label{eq:system}
	\begin{align}
		z_t(x, t) = \left[\theta(x) z_x(x,t)\right]_x &+ \lambda(x) z(x, t), \label{eq:PDE}
	\end{align}
the Robin BCs
\begin{align}
	-b_0 z_x(0,t) + a_0 z(0,t) &= u_0(t) + \psi_0(t), \label{eq:BC0} \\
	b_1 z_x(1,t) + a_1 z(1, t) &= u_1(t) + \psi_1(t), \label{eq:BC1}
\end{align}
and the initial condition $z(x, 0) = z_0(x) \in L_2(0, 1)$.
\end{subequations}
The functions $\theta(x)$ and $\lambda(x)$ denote the spatially varying diffusion and reaction coefficients, respectively. The diffusion coefficient $\theta(x)$ is assumed to be of class $C^1$. Both $\theta(x)$ and $\lambda(x)$ are unknown and bounded, as formulated in the following
\begin{assumption} \label{ass:theta_lambda}
	There exist constants $\theta_\mathrm{m}$, $\theta_\mathrm{M}$, $\lambda_\mathrm{m}$ and $\lambda_\mathrm{M}$ such that the next inequalities hold $\forall x \in [0, 1]$
	\begin{eqnarray}
		0 &<& \theta_\mathrm{m} \le \theta(x) \le \theta_\mathrm{M}, \quad \lambda_\mathrm{m} \le \lambda(x) \le \lambda_\mathrm{M},\\
		\lambda_\mathrm{M} &<& \frac{\pi^2}{4}\theta_\mathrm{m}.\label{eq:lambda_ass}
	\end{eqnarray}
\end{assumption}
\vspace{-3mm}
Similarly, the constants $a_i$, $b_i$ of the Robin BCs \cref{eq:BC0,eq:BC1} are unknown and subject to the next
\begin{assumption} \label{ass:aibi}
	There exist constants $\am{i}$, $\aM{i}$ and $\bM{i}$ ($i \in \{0,1\}$) such that
	\begin{align}
		\am{i} \le a_i \le \aM{i}, \quad 0 < b_i \le \bM{i}.
	\end{align}
\end{assumption}
\vspace{-3mm}
If $\lambda_M \leq 0$ then all the constants involved in the \cref{ass:theta_lambda} and \cref{ass:aibi} (except $\am{i}$) can be unknown. Note that, in this favourable scenario, relation \eqref{eq:lambda_ass} is always fulfilled by construction. If, on the contrary, $\lM > 0$, then the scenario is more challenging since the reaction term has a destabilizing effect. In this case, $\lM$, $\thetam$ and  $\bM{i}$ must be known in advance. \\
Signals $u_0(t)$ and $u_1(t)$ represent the manipulable boundary control inputs, applied through the Robin BCs \cref{eq:BC0,eq:BC1}. The system is further affected by the matched boundary disturbances $\psi_0(t)$ and $\psi_1(t)$ which are both assumed to be uniformly bounded according to
\begin{assumption} \label{ass:disturbances}
	There exist unknown constants $\Phi_0$ and $\Phi_1$ such that
	\begin{align}
		|\psi_i(t)| \le \Phi_i, \quad \forall t \ge 0, \quad {i \in \{0, 1\}}. \label{eq:boundpsi}
	\end{align}
     \end{assumption} 
     \vspace{-3mm}
Employing the monodirectional adaptation, to be introduced later on, additionally involves a restriction on the time derivative of the disturbance signals, specified by the next
      \begin{assumption} \label{ass:disturbancesderivative}
       There exist unknown constants $\Phid{0}$ and $\Phid{1}$ such that
        \begin{align}
            \left|\dv{t}\psi_i(t)\right| &\le \Phid{i}, \quad \forall t \ge 0, \quad i \in \{0, 1\}.  \label{eq:boundpsidot}
        \end{align}
   \end{assumption} 
   \vspace{-3mm}

The control goal is to design suitable control signals $u_0(t)$ and $u_1(t)$ capable of steering the $L_2$-norm of the state $z(x,t)$ either to zero, or to a vicinity around zero. This should happen despite the presence of the unknown disturbances $\psi_0(t)$ and $\psi_1(t)$ with unknown bounds and unknown system parameters. In the next section, control laws are proposed that achieve these goals.

\section{Controller synthesis} \label{sec:controller}
A proportional and discontinuous feedback of the form
\begin{subequations} \label{eq:controller}
\begin{align}
	u_i(t) &= -\propgain_i z(x_i, t) - M_i(t) \sign{z(x_i, t)} \label{eq:control} \\
    \dot{M}_i(t) &= -\alpha_i M_i(t) + \gamma_i |z(x_i, t)| \label{eq:control_adapt}
\end{align}
\end{subequations}
is proposed, where $i \in \{0, 1\}$ and $x_0 = 0$, $x_1 = 1$
denote the two boundaries of the spatial domain.

The adaptive switching gains $M_i(t)$ evolve according to the adaptation laws \cref{eq:control_adapt} where the initial values $M_i(0) = M_{0,i}$ satisfy
\begin{align}
	M_{0,i} \ge 0, \quad \forall i \in \{0,1\}. \label{eq:tuning_M0}
\end{align}
The constant gains $\gamma_i$ and $\alpha_i$ are tuned according to
\begin{align}
    \gamma_i & > 0, \label{eq:tuning_gamma} \\
    \alpha_i &\ge 0. \label{eq:tuning_alpha}
\end{align}
The parameters $\alpha_i$ determine whether the adaptation is monodirectional or bidirectional. By choosing ${\alpha_i = 0}$, the right-hand side of \cref{eq:control_adapt} is non-negative which implies monodirectional adaptation, that is, the gains $M_i(t)$ cannot decrease. The choice $\alpha_i > 0$ allows the right-hand side of \cref{eq:control_adapt} to become negative which leads to possibly decreasing gains $M_i(t)$, hence, this case is referred to as bidirectional adaptation. In the upcoming stability analysis, the cases $\alpha_0, \alpha_1 > 0$ and $\alpha_0 = \alpha_1 = 0$ are considered. That is, either both adaptation laws are bidirectional or both are monodirectional. 

To achieve the control goal it is later shown that in the bidirectional adaptation case $\alpha_0, \alpha_1 > 0$, the proportional gains $k_0$ and $k_1$ are required to satisfy
\begin{subequations} \label{eq:tuning_k_bi}
\begin{align} 
	k_0 &> \frac{\overline{\lM} \pi^2}{\thetam \pi^2 - 4 \overline{\lM}}\bM{0} -\am{0} \label{eq:tuning_k0_bi}, \\
        k_1 &\ge -\am{1}, \label{eq:tuning_k1_bi}
\end{align}
\end{subequations}
where $\overline{\lM} \coloneqq \max\{0, \lM\}$.
For the monodirectional adaptation case, i.e. $\alpha_0 = \alpha_1 = 0$, the parameter conditions to be imposed are
\begin{subequations}\label{eq:tuning_k_mono}
\begin{align}
    k_0 &\ge \frac{\overline{\lM} \pi^2}{\thetam \pi^2 - 4 \overline{\lM}}\bM{0} -\am{0} +\gamma_0, \label{eq:tuning_k0_mono}\\
    k_1  &\ge  - \am{1} + \gamma_1, \label{eq:tuning_k1_mono}
\end{align}
\end{subequations}

\subsection{Well posedness of the closed-loop system} 

The proposed control input \cref{eq:controller} undergoes discontinuities in the manifolds $z(0,t) = 0$ and $z(1,t) = 0$. Similar to \cite[Definition 1]{Pisano2017} the meaning of the closed-loop system \cref{eq:system}, driven by \cref{eq:controller} is adopted in the weak sense beyond the discontinuity manifold, otherwise, it is viewed in the Filippov sense. In addition to Reference \cite{Pisano2017}, the interested reader may also refer to Reference \cite{Orlov2022book} for more details on weak and Filippov (sliding mode) solutions in the PDE setting. Since the above closed-loop system is of class $C^1$ beyond its discontinuity manifold, it possesses a unique local weak solution once initialized with $z_0(x)$ such that $z_0(0) \neq 0$ and $z_0(1) \neq 0$, \cite[Theorem 23.2]{Krasnoselskii1976book}. If a sliding mode occurs on, e.g., the discontinuity manifold $z(0, t) = 0$ then it is governed by the same PDE \eqref{eq:PDE} subject to the mixed-type boundary conditions formed by  $z(0, t) = 0$ and the Robin-type boundary condition \eqref{eq:BC1}  that remains in force, which is of class $C^1$ and thus well-posed. Similar considerations can be done to analyze the sliding mode solutions along the discontinuity manifold $z(1, t) = 0$ and along their intersection $z(0, t)=z(1, t) = 0$.

\subsection{Main result}
A useful inequality to be used throughout the upcoming stability analyses, is presented in the next Lemma.
\begin{lemma} \label{thm:lemmaPoincare}
    Let $f \in H^1(0,1)$ and define $ \bar{f} = \int_0^1 f(x)~\dd x$. Then it holds that
    \begin{align}
        -\frac{\pi^2}{4} \ltwonorm{f(\cdot)}^2 + \frac{\pi^2}{2} f(0)\bar{f} - \frac{\pi^2}{4}f^2(0) &\ge -\ltwonorm{f'(\cdot)}^2 \label{eq:poincare_use}
    \end{align}
\end{lemma}
\begin{pf}
Consider the Poincaré-type inequality (see, e.g. \cite{Smyshlyaev2010})
    \begin{align}
        \ltwonorm{f(\cdot) - f(0)}^2 &\le \frac{4}{\pi^2} \ltwonorm{f'(\cdot)}^2. \label{eq:poincare_new}
    \end{align}
Expanding the left-hand side of \cref{eq:poincare_new} to
    \begin{align}
        \ltwonorm{f(\cdot) - f(0)}^2 &= \int_0^1 \left(f(x) - f(0)\right)^2~\dd x \nonumber \\
        &=  \ltwonorm{f(\cdot)}^2 - 2 f(0)\bar{f} + f^2(0)
    \end{align}
    leads to \cref{eq:poincare_use} after rearranging terms. $\hfill \square$
\end{pf}
The following Theorem \ref{th:main} constitutes the main theoretical result of the present paper and presents the Lyapunov-based stability analysis of the closed-loop system. Two cases (bidirectional and monodirectional adaptation) are distinguished. In the former case a stability result as in the notion of globally uniformly ultimately bounded solutions (\cite[Definition 4.6]{Khalil2002}) for the $L_2$-space is achieved, whereas in the latter case it is shown that the origin is globally asymptotically stable in the $L_2$-sense.
\begin{thm} \label{th:main}
	Consider System \cref{eq:system} along with the adaptive boundary control laws \cref{eq:controller}, and let \cref{ass:theta_lambda,ass:aibi,ass:disturbances} be fulfilled. Let the control parameters be tuned according to \eqref{eq:tuning_M0} and \eqref{eq:tuning_gamma}. The following statements hold:
    \begin{enumerate}[label=(\roman*)]
        \item \label{item:bidir} \textbf{Bidirectional adaptation}. Let $\alpha_0,\alpha_1 > 0$ and  \cref{eq:tuning_k_bi} be fulfilled. Then, there exist constants $T$ and $\mathcal{B}$ such that
        \begin{align}
            \ltwonorm{z(\cdot, t)} \le \mathcal{B}, \quad \forall t \ge T,~ \forall z_0(x). \label{eq:biresult}
        \end{align}
        \item \label{item:monodir} \textbf{Monodirectional adaptation}. Let $\alpha_0 =\alpha_1 = 0$, \cref{ass:disturbancesderivative} be in force, and  \cref{eq:tuning_k_mono} be fulfilled. Then, the zero solution 
	$z^*(x, t) = 0$ is globally asymptotically stable in the $L_2(0,1)$-sense, that is
	\begin{align}
	\lim_{t\to\infty}\ltwonorm{z(\cdot, t) - z^*(\cdot, t)} = 0, \quad\forall z_0(x).
	\end{align}
    \end{enumerate}  
\end{thm}

\begin{pf}
	The proof in the case of bidirectional adaptation differs from the proof in case of monodirectional adaptation. However, both proofs share a common part which is presented first. The result of the common part is an inequality regarding the time derivative of a Lyapunov function candidate. This result is then manipulated and analyzed further individually for the two distinct cases \labelcref{item:bidir,item:monodir}. For the closed-loop system in question, consider the Lyapunov function candidate
	\begin{align} \label{eq:V}
		V(t) &= V_1(t) + V_2(t)
	\end{align}
	where
	\begin{align}
		V_1(t) &= \frac{1}{2} \ltwonorm{z(\cdot, t)}^2 = \frac{1}{2} \int_0^1 z^2(x, t)~\dd x \label{eq:V1}
	\end{align}
	and
	\begin{align} 
		V_2(t) &= \sum_{i=0}^1 \frac{\theta(x_i)}{2 b_i \gamma_i} \left(M_i(t) - \Phi_i\right)^2. \label{eq:V2}
	\end{align}
	Strictly speaking $V_1(t)$ is a functional but for simplicity it is referred to as a function. Furthermore, ${V_1(z(\cdot, t)) = V_1(t)}$ is written for this and, analogously, for other functions.
	The time derivative of $V_1(t)$ calculates as
	\begin{align}
		\dot{V}_1(t) &= \int_0^1z(x,t)z_t(x,t)~\dd x
	\end{align}
	which, when inserting PDE \cref{eq:PDE}, yields
	\begin{align}
\dot{V}_1(t) &= \int_0^1z(x,t) \left[\theta(x) z_x(x,t)\right]_x ~\dd x \alignbreak + \int_0^1\lambda(x) z^2(x, t) ~\dd x. \label{eq:intparts}
	\end{align}
	Applying integration by parts to the first integral in \cref{eq:intparts} gives
	\begin{align}
		\dot{V}_1(t) &= \theta(x) z_x(x,t) z(x,t)\big|_0^1 + \kappa(t) \label{eq:insertBC}
	\end{align}
	where
	\begin{align}
		\kappa(t) \coloneqq -\int_0^1 \theta(x) z^2_x(x,t) ~\dd x + \int_0^1 \lambda(x) z^2(x, t) ~\dd x. \label{eq:kappadef}
	\end{align}
	The BCs \cref{eq:BC0,eq:BC1} are rearranged and inserted into \cref{eq:insertBC}, resulting in
	\begin{align}
		\dot{V}_1(t) &=\sum_{i=0}^1	\frac{\theta(x_i)}{b_i}\left[u_i(t) + \psi_i(t) - a_i z(x_i, t)\right]z(x_i, t) \alignbreak
		+\kappa(t). \label{eq:insertu}
	\end{align}
	By substituting control laws \cref{eq:control} into the right-hand side of \cref{eq:insertu}
    one ends up with
	\begin{align}
		\dot{V}_1(t) &= \kappa(t) + \sum_{i=0}^1\frac{\theta(x_i)}{b_i}\big[-\ka{i} z^2(x_i, t) \alignbreak- M_i(t) |z(x_i, t)| + \psi_i(t)z(x_i, t) \big] \label{eq:V1dot}
	\end{align}
        where
        \begin{align}
            \ka{i} \coloneqq k_i + a_i. \label{eq:ka}
        \end{align}
	Differentiating \eqref{eq:V2}, and considering \cref{eq:control_adapt}, yields
	\begin{align}
		\dot{V}_2(t) &= \sum_{i=0}^1 \frac{\theta(x_i)}{b_i} \Big[M_i(t) |z(x_i, t)| - \Phi_i |z(x_i, t)| \alignbreak+ \frac{\alpha_i}{\gamma_i} \Phi_i M_i(t) - \frac{\alpha_i}{\gamma_i} M_i^2(t)\Big]. \label{eq:V2dot}
	\end{align}
	Combining \cref{eq:V1dot,eq:V2dot} yields
	\begin{align}
		\dot{V}(t) &= \dot{V}_1(t) + \dot{V}_2(t) \\
		&= \kappa(t) + \sum_{i=0}^1 \frac{\theta(x_i)}{b_i}\big[-\ka{i} z^2(x_i, t) + \psi_i(t)z(x_i, t) \alignbreak- \Phi_i |z(x_i, t)|+ \frac{\alpha_i}{\gamma_i} \Phi_i M_i(t) - \frac{\alpha_i}{\gamma_i} M_i^2(t)\big]. \label{eq:Vdoteq}
        \end{align}
        Two terms in \cref{eq:Vdoteq} are estimated by
        \begin{align}
            \psi_i(t)&z(x_i, t) - \Phi_i |z(x_i, t)| \\ \nonumber
            &\quad\le (|\psi_i(t)| - \Phi_i) |z(x_i, t)| \le 0
        \end{align}
        which holds due to \cref{ass:disturbances}.
By considering \cref{ass:theta_lambda} in \cref{eq:kappadef}, inequality \cref{eq:Vdoteq} is further manipulated to
        \begin{align}
        \dot{V}(t)	&\le -\thetam \ltwonorm{z_x(\cdot, t)}^2 + \lM \ltwonorm{z(\cdot, t)}^2 \alignbreak - \sum_{i=0}^1 \frac{\theta(x_i)}{b_i}\ka{i} z^2(x_i, t) \alignbreak+ \sum_{i=0}^1 \alpha_i \frac{\theta(x_i)}{b_i \gamma_i} \left[ \Phi_i M_i(t) -  M_i^2(t) \right]. \label{eq:Vdot}
	\end{align}
    As in \cite{Roy2020}, the estimation
    \begin{align}
        \Phi_i M_i(t) - M_i^2(t) &= -\frac{1}{2}\left(M_i(t) - \Phi_i\right)^2 - \frac{1}{2}M_i^2(t) + \frac{1}{2} \Phi_i^2 \nonumber \\
	&\le -\frac{1}{2}\left(M_i(t) - \Phi_i\right)^2 + \frac{1}{2} \Phi_i^2 
    \end{align}
    is considered, by means of which the next relation
    \begin{align}
        \dot{V}(t) &\le - \thetam \ltwonorm{z_x(\cdot, t)}^2 + \lM \ltwonorm{z(\cdot, t)}^2 \alignbreak - \sum_{i=0}^1\frac{\theta(x_i)}{b_i}\ka{i} z^2(x_i,t) - \sum_{i=0}^1 \alpha_i\frac{\theta(x_i)}{2b_i\gamma_i}  \left(M_i(t) - \Phi_i\right)^2 \alignbreak + \sum_{i=0}^1\alpha_i \frac{\theta(x_i)}{2b_i\gamma_i} \Phi_i^2 \label{eq:drop_k1}
    \end{align}
    is obtained. When considering \cref{ass:theta_lambda,ass:aibi}, parameter conditions \cref{eq:tuning_k_bi} and \cref{eq:tuning_k_mono} both imply that
    \begin{align}
        \ka{i} \ge 0, \quad\forall i \in \{0, 1\}. \label{eq:ka_pos}
    \end{align}
    Together with \cref{ass:theta_lambda} this allows dropping the term corresponding to $\bar{k}_1$ in \cref{eq:drop_k1}. Only the stabilizing part corresponding to $\bar{k}_0$ is needed for the ongoing development. With the definition of $V_2$ in \cref{eq:V2} it therefore holds that
    \begin{align}
        \dot{V}(t) &\le - \thetam \ltwonorm{z_x(\cdot, t)}^2 + \lM \ltwonorm{z(\cdot, t)}^2 \alignbreak - \frac{\thetam}{b_0}\ka{0} z^2(0,t) - \min_{i}(\alpha_i) V_2(t) \alignbreak + 
        \sum_{i=0}^1\alpha_i \frac{\theta(x_i)}{2b_i\gamma_i} \Phi_i^2.
    \end{align}
    As in \cite{Pisano2017} the solutions $z(\cdot, t)$ are considered to lie in $H^1(0,1)$, hence, \cref{thm:lemmaPoincare} is applied to obtain that
    \begin{align}
        \dot{V}(t) &\le \left(\lM -\thetam\frac{\pi^2}{4} \right)\ltwonorm{z(\cdot, t)}^2 + \thetam\frac{\pi^2}{2} z(0, t)\bar{z}(t) \alignbreak- \thetam \left(\frac{\pi^2}{4} + \frac{\ka{0}}{b_0}\right)z^2(0, t) \alignbreak -\min_{i}(\alpha_i) V_2(t) + \sum_{i=0}^1 \alpha_i \frac{\theta(x_i)}{2b_i\gamma_i} \Phi_i^2 \label{eq:Vdot_after_poincare}
    \end{align}
    where $\bar{z}(t) \coloneqq \int_0^1 z(x, t)~\dd x$.
Due to \cref{eq:ka_pos,ass:aibi} it holds that
    \begin{align}
        \frac{\ka{0}}{b_0} + \frac{\pi^2}{4} > 0 \label{eq:posterm}
    \end{align}
which allows completing the square, thereby yielding
    \begin{align}
        \dot{V}(t) &\le\left(\lM -\thetam\frac{\pi^2}{4} \right)\ltwonorm{z(\cdot, t)}^2 \alignbreak- \thetam \left(\sqrt{\frac{\ka{0}}{b_0} + \frac{\pi^2}{4}}z(0, t) - \frac{\pi^2 \bar{z}(t)}{4 \sqrt{\frac{\ka{0}}{b_0} + \frac{\pi^2}{4}}} \right)^2 \alignbreak+ \thetam \frac{\pi^4 \bar{z}^2(t)}{16 \left(\frac{\ka{0}}{b_0} + \frac{\pi^2}{4}\right)} -\min_{i}(\alpha_i) V_2(t) \alignbreak+ \sum_{i=0}^1 \alpha_i \frac{\theta(x_i)}{2b_i\gamma_i} \Phi_i^2 \\
        & \le\left(\lM -\thetam\frac{\pi^2}{4} \right)\ltwonorm{z(\cdot, t)}^2 \alignbreak+ \thetam \frac{\pi^4 \bar{z}^2(t)}{16 \left(\frac{\ka{0}}{b_0} + \frac{\pi^2}{4}\right)} -\min_{i}(\alpha_i) V_2(t) \alignbreak+ \sum_{i=0}^1 \alpha_i \frac{\theta(x_i)}{2b_i\gamma_i} \Phi_i^2 \label{eq:Vdothoelder}
    \end{align}
       It is now desired to combine the first two terms in \cref{eq:Vdothoelder}. It holds that
    \begin{align}
        \bar{z}(t) &\le \int_0^1|z(x,t)|~\dd x \eqqcolon \lonenorm{z(\cdot, t)}
    \end{align}
    while Hölder's inequality (see, e.g., \cite[Theorem 4.6]{Brezis2010}) implies that $\lonenorm{z(\cdot, t)} \le \ltwonorm{z(\cdot, t)}$.
Thus
    \begin{align}
        \bar{z}^2(t) &\le \ltwonorm{z(\cdot, t)}^2. \label{eq:dc_less_norm}
    \end{align}
Applying \cref{eq:dc_less_norm} in \cref{eq:Vdothoelder} yields
    \begin{align}
    	\dot{V}(t) &\le -\beta \ltwonorm{z(\cdot, t)}^2 -\min_{i}(\alpha_i) V_2(t) + \sum_{i=0}^1 \alpha_i \frac{\theta(x_i)}{2b_i\gamma_i} \Phi_i^2 \label{eq:Vdot2}
    \end{align}
    where
    \begin{align}
    	\beta \coloneqq \thetam\frac{\pi^2}{4} -\lM - \thetam \frac{\pi^4}{16 \left(\frac{\ka{0}}{b_0} + \frac{\pi^2}{4}\right)}. \label{eq:betadef}
    \end{align}
    It is desired to tune the control parameter $\bar{k}_0$ such that
    \begin{align}
        \beta > 0 \label{eq:betapos}.
    \end{align}
    To this end, inequality \cref{eq:betapos} is written as
    \begin{align}
        4 \thetam \pi^2 \left( \frac{\ka{0}}{b_0} + \frac{\pi^2}{4}\right) - 16 \lM \left( \frac{\ka{0}}{b_0} + \frac{\pi^2}{4}\right) - \thetam \pi^4 > 0  \label{eq:manipu01}
    \end{align}
    which holds because of \cref{eq:posterm}. Manipulating \eqref{eq:manipu01} yields
    \begin{align}
        \frac{\ka{0} }{b_0}\left( \thetam \pi^2 - 4 \lM \right) - \lM \pi^2 > 0 
    \end{align}
    which is rearranged to
    \begin{align}
        k_0 > \frac{\lM \pi^2}{\thetam \pi^2 - 4 \lM} b_0 - a_0 \label{eq:ineq_k0}
    \end{align}
    by considering \cref{eq:ka}.
    Parameter conditions \cref{eq:tuning_k0_bi,eq:tuning_k0_mono} are both sufficient for \cref{eq:ineq_k0}, hence, ${\beta > 0}$.
    
    This concludes the common part of the proof for the bidirectional and monodirectional adaptation cases \labelcref{item:bidir,item:monodir}. Inequality \cref{eq:Vdot2} is now further manipulated for the two cases, starting with the bidirectional adaptation case \labelcref{item:bidir}.
    
    \subsection*{\labelcref{item:bidir} Bidirectional adaptation: $\alpha_0, \alpha_1 > 0$}
    In the remainder, $V(t)$ is identified in the right-hand side of $\dot{V}(t)$ with which the bound $\mathcal{B}$ from \cref{eq:biresult} is derived. \\
    Inserting the definition of $V_1(t)$ from \cref{eq:V1} into \cref{eq:Vdot2} yields
\begin{align}
	\dot{V}(t) &\le -2\beta V_1(t) -\min_{i}(\alpha_i) V_2(t) + \sum_{i=0}^1 \alpha_i \frac{\theta(x_i)}{2b_i\gamma_i} \Phi_i^2. \label{eq:vdotv1insert}
\end{align}
The constant $\beta$ from \cref{eq:betadef} can be estimated by
\begin{align}
    \beta > \beta_\text{min} &\coloneqq \thetam \frac{\pi^2}{4} - \lM -\thetam \frac{\pi^4}{16 \left( \frac{k_0 + \am{0}}{\bM{0}} + \frac{\pi^2}{4} \right)} \label{eq:betamindef}
\end{align}
by considering \cref{ass:aibi}. Applying \cref{eq:betamindef} to \cref{eq:vdotv1insert} results in
\begin{align}
    \dot{V}(t) &\le -2\beta_\text{min} V_1(t) -\min_{i}(\alpha_i) V_2(t) + \sum_{i=0}^1 \alpha_i \frac{\theta(x_i)}{2b_i\gamma_i} \Phi_i^2. \nonumber
\end{align}
A constant $\rho$ is defined as
\begin{align}
	\rho &\coloneqq \operatorname{min}\left\{2\beta_\text{min}, \alpha_0, \alpha_1\right\} > 0 \label{eq:rhodef}
\end{align}
which is positive since $\alpha_i > 0$ $\forall i \in \{0, 1\}$ in the bidirectional case. This definition enables writing
\begin{align}
	\dot{V}(t) &\le -\rho \left(V_1(t) + V_2(t)\right) + \sum_{i=0}^1 \alpha_i \frac{\theta(x_i)}{2b_i\gamma_i} \Phi_i^2
\end{align}
which can be simplified to
\begin{align}
	\dot{V}(t) &\le -\rho V(t) + \eta \label{eq:Vdot1}
\end{align}
by taking into account \cref{eq:V,ass:theta_lambda,ass:aibi} and defining
\begin{align}
	\eta \coloneqq \sum_{i=0}^1 \alpha_i \frac{\theta_\text{M}}{2\bm{i}\gamma_i} \Phi_i^2.
\end{align}
In order to define a set which is reached in finite time as in \cite{Roy2020}, the variable $\varepsilon$ is introduced by $0 < \varepsilon < \rho$
with which \cref{eq:Vdot1} is expressed as
\begin{align}
	\dot{V}(t) &\le -\varepsilon V(t) \underbrace{- \left(\rho - \varepsilon\right)V(t) + \eta}_{\chi}. \label{eq:Vdot_chi}
\end{align}
The last two terms in \cref{eq:Vdot_chi} are collected in a new variable $\chi \coloneqq - \left(\rho - \varepsilon\right)V(t) + \eta$.
It can be seen that $\chi \le 0$ is equivalent to
\begin{align}
    V(t) \ge \frac{\eta}{\rho - \varepsilon}. \label{eq:Vbound}
\end{align}
Hence, whenever \cref{eq:Vbound} holds, inequality \cref{eq:Vdot_chi} is rendered to $\dot{V}(t) \le -\varepsilon V(t)$
and by comparison principle it holds that $V(t) \le V(0) e^{-\varepsilon t}$.
Thus, in finite time $V(t)$ reaches the set defined by $V(t) \le \frac{\eta}{\rho - \varepsilon}$.
Furthermore, since $V_1(t) = \frac{1}{2}\ltwonorm{z(\cdot, t)}^2 \le V(t)$, the state norm enters in finite time the set defined by $\ltwonorm{z(\cdot, t)} \le \mathcal{B}$ where the bound $\mathcal{B}$ from \cref{eq:biresult} is given by
\begin{align}
    \mathcal{B} =  \sqrt{\frac{2\eta}{\rho - \varepsilon}}. \label{eq:boundB}
\end{align}
This concludes the proof for the bidirectional adaptation case.

    \subsection*{\labelcref{item:monodir} Monodirectional adaptation: $\alpha_0 = \alpha_1 = 0$}
    In the monodirectional adaptation case a term corresponding to $V_2(t)$ is not found in the right-hand side of $\dot{V}(t)$. Hence, also $V(t)$ cannot be found in the right-hand side of $\dot{V}(t)$. Therefore, a different argumentation is used for obtaining a stability result. The approach taken in \cite{Mayr2024} is followed where Barbalat's Lemma is employed to show asymptotic convergence of the state norm to zero. Before this step, some preliminary results about the state norm $\ltwonorm{z(\cdot, t)}$, the adaptive gains $M_i(t)$ and other terms are obtained. \\ The analysis starts by considering $\alpha_0 = \alpha_1 = 0$ in inequality \cref{eq:Vdot2}, which is therefore reduced to
    \begin{align}
        \dot{V}(t) &\le -\beta \ltwonorm{z(\cdot, t)}^2 \le 0. \label{eq:Vdotle0}
    \end{align}
    Hence, $V(t)$ is nonincreasing, which implies
	\begin{align}
		0 \le V(t) \le V(0) \;\;\; \forall t \ge 0. \label{eq:Vbnd}
	\end{align}
	From \cref{eq:Vbnd,eq:V,eq:V1,eq:V2}, and taking into account that $V_i(t) \le V(t) \le V(0)$, $\forall i=1,2$,
	one can derive the following uniform upper bounds for $\ltwonorm{z(\cdot,t)}$ and $M_i(t)$
	\begin{align}
		\ltwonorm{z(\cdot,t)} \le \sqrt{2 V(0)}, \label{eq:bndNormz} \\
		M_i(t) \le  \Gamma_i \coloneqq \Phi_i + \sqrt{\frac{2 b_i \gamma_i}{\theta(x_i)} V(0)} , \quad i \in \{0, 1\} \label{eq:bndM}
	\end{align}

	\subsection*{Uniform boundedness of $\int_0^1 \theta(x) z^2_x(x,t) ~\dd x$ and $|z(x_i, t)|$}
	The new Lyapunov candidate function
	\begin{align}
		W(t) &= \sum_{i=1}^8 V_i(t)
	\end{align}
	is introduced with
	\begin{align}
		V_3(t) &= \frac{1}{2} \int_0^1 \theta(x) z_x^2(x, t)~\dd x, \label{eq:V3} \\
		V_4(t) &= \sum_{i=0}^1 \frac{\theta(x_i)}{b_i} M_i(t) |z(x_i, t)|, \label{eq:V4} \\
		V_5(t) &= \sum_{i=0}^1 \frac{\theta(x_i)}{b_i} \left[\Phi_i |z(x_i, t)| - \psi_i(t) z(x_i, t)\right], \label{eq:V5} \\
		V_6(t) &= \sum_{i=0}^1 \frac{\theta(x_i)}{b_i} \frac{\Phid{i}}{\gamma_i} \left(\Gamma_i - M_i(t)\right), \label{eq:V6} \\
		V_7(t) &= \sum_{i=0}^1 \frac{\theta(x_i)}{2b_i} \left( \sqrt{\ka{i}} |z(x_i, t)| - \frac{1}{\sqrt{\ka{i}}} \Phi_i \right)^2, \label{eq:V7} \\
		V_8(t) &= |\lM| V(t) - \frac{1}{2} \int_0^1 \lambda(x)z^2(x,t)~\dd x. \label{eq:V8}
	\end{align}
	The non-negativeness of $V_5(t)$ derives from the inequalities
	\begin{equation}
		\psi_i(t) z(x_i, t) \le |\psi_i(t) z(x_i, t)| \le \Phi_i |z(x_i, t)|,
	\end{equation}
	whereas the non-negativeness of $V_6(t)$ is due to the previously proven relation \eqref{eq:bndM}, namely the existence of the uniform upper bounds $\Gamma_i$ for the adaptive gains $M_i(t)$.
    The term $V_8(t)$ is non-negative since
    \begin{align}
        \frac{1}{2}\int_0^1 \lambda(x) z^2(x,t) ~ \dd x \le \frac{|\lM|}{2} \ltwonorm{z(\cdot,t)}^2 \le |\lM| V(t)
    \end{align}
    which holds due to \cref{ass:theta_lambda} and the definition of $V(t)$ in \cref{eq:V,eq:V1,eq:V2}.
	The time derivatives of the newly introduced functions \cref{eq:V3,eq:V4,eq:V5,eq:V6,eq:V7,eq:V8} along the closed-loop system's solutions are now calculated, starting with $\dot{V}_3(t)$ for which integration by parts is applied. This yields
	\begin{align}
		\dot{V}_3(t) &= \int_0^1 \theta(x) z_x(x,t) z_{xt}(x,t) ~ \dd x \nonumber \\
		&= z_t(x,t) \theta(x) z_x(x,t)\big|_0^1 \alignbreak
		- \int_0^1 z_t(x,t) \underbrace{\left[\theta(x) z_x(x,t)\right]_x}_{z_t(x,t) - \lambda(x) z(x, t)} ~\dd x \label{eq:V3dot01}
	\end{align}
	Substituting the BCs \cref{eq:BC0,eq:BC1} and the boundary control law \cref{eq:control} with \cref{eq:ka} into \cref{eq:V3dot01} gives
\begin{align}
		\dot{V}_3(t) &= -\ltwonorm{z_t(\cdot, t)}^2 + \int_0^1\lambda(x) z(x,t) z_t(x,t) ~ \dd x \alignbreak
		+ \sum_{i=0}^1 \frac{\theta(x_i)}{b_i} [-\ka{i} z(x_i, t)z_t(x_i, t) \alignbreak
		- M_i(t) \sign{z(x_i, t)}z_t(x_i, t) + \psi_i(t)z_t(x_i, t)]. \label{eq:V3dot}
	\end{align}
	Evaluating the time derivatives of the remaining functions ${V}_4(t)$ to ${V}_8(t)$  yields
	\begin{align}
\dot{V}_4(t) &= \sum_{i=0}^1 \frac{\theta(x_i)}{b_i} \big[M_i(t) z_t(x_i, t)\sign{z(x_i, t)} \alignbreak
		+ \gamma_i z^2(x_i, t)\big], \label{eq:V4dot} \\
		\dot{V}_5(t) &= \sum_{i=0}^1 \frac{\theta(x_i)}{b_i} \big[\Phi_i z_t(x_i, t) \sign{z(x_i, t)} \alignbreak - \dot{\psi}_i(t) z(x_i, t) - \psi_i(t) z_t(x_i, t)\big], \label{eq:V5dot} \\
		\dot{V}_6(t) &= -\sum_{i=0}^1 \frac{\theta(x_i)}{b_i} \Phid{i} |z(x_i, t)|, \label{eq:V6dot} \\
\dot{V}_7(t) &= \sum_{i=0}^1 \frac{\theta(x_i)}{b_i} \big[\ka{i} z(x_i, t) z_t(x_i, t) \alignbreak - \Phi_i z_t(x_i, t)\sign{z(x_i, t)} \big], \label{eq:V7dot} \\
		\dot{V}_8(t) &= |\lM| \dot{V}(t) - \int_0^1 \lambda(x)z(x, t)z_t(x, t)~ \dd x \label{eq:V8dot}
	\end{align}
    with $\dot{V}(t)$ in \cref{eq:V8dot} from \cref{eq:Vdoteq}.
	Combining \cref{eq:Vdoteq} and \cref{eq:V3dot,eq:V4dot,eq:V5dot,eq:V6dot,eq:V7dot,eq:V8dot}, and reordering, leads to
	\begin{align}
		\dot{W}(t) &= \sum_{i = 1}^8 \dot{V}_i(t)= \kappa(t) -\ltwonorm{z_t(\cdot, t)}^2 + |\lM| \dot{V}(t) \alignbreak
		- \sum_{i=0}^1 \frac{\theta(x_i)}{b_i}\Big[(\ka{i} - \gamma_i) z^2(x_i, t) \alignbreak
		+\left[\psi_i(t) - \dot{\psi}_i(t)\right] z(x_i, t)  - \left[\Phi_i + \Phid{i}\right] |z(x_i, t)| \Big]. \label{eq:Wdot01}
	\end{align}
	The sign-indefinite term $\left[\psi_i(t) - \dot{\psi}_i(t)\right] z(x_i, t)$ in the right-hand side of \cref{eq:Wdot01} can be estimated as
	\begin{align}
		\left|\left[\psi_i(t) - \dot{\psi}_i(t)\right] z(x_i, t)\right| \le \left[\Phi_i + \Phid{i}\right] |z(x_i, t)| \label{eq:estbndWdot}
	\end{align}
	by virtue of \cref{eq:boundpsi,eq:boundpsidot}.    
	Thus, by virtue of \cref{eq:Vdotle0,eq:estbndWdot} one can further manipulate the right-hand side of \eqref{eq:Wdot01} to get
	\begin{align}
		\dot{W}(t) \le \kappa(t) - \sum_{i=0}^1 \frac{\theta(x_i)}{b_i}\Big[(\ka{i} - \gamma_i) z^2(x_i, t)\Big].
	\end{align}
    By considering \cref{eq:kappadef,ass:theta_lambda} it is obtained that
    \begin{align}
        \dot{W}(t) &\le -\thetam \ltwonorm{z_x(\cdot, t)}^2 + \lM \ltwonorm{z(\cdot, t)}^2 \alignbreak- \sum_{i=0}^1 \frac{\theta(x_i)}{b_i}\Big[(\ka{i} - \gamma_i) z^2(x_i, t)\Big].
    \end{align}
    Parameter conditions \cref{eq:tuning_k_mono} imply that
    \begin{align}
        \ka{i} - \gamma_i \ge 0,\quad \forall i \in \{0, 1\}. \label{eq:kagammapos}
    \end{align}
    This is used together with \cref{thm:lemmaPoincare,ass:theta_lambda} to get
    \begin{align}
        \dot{W}(t) &\le \left(\lM -\thetam\frac{\pi^2}{4} \right)\ltwonorm{z(\cdot, t)}^2 + \thetam\frac{\pi^2}{2} z(0, t)\bar{z}(t) \alignbreak- \thetam \left(\frac{\ka{0} - \gamma_0}{b_0} + \frac{\pi^2}{4}\right)z^2(0, t). \label{eq:Wdot_after_poincare}
    \end{align}
    Note that the right-hand side of \cref{eq:Wdot_after_poincare} is the same expression as in \cref{eq:Vdot_after_poincare} with $\alpha_0 = \alpha_1 = 0$ and the substitution $\ka{0} \leftarrow \ka{0} - \gamma_0$. Because of \cref{eq:kagammapos} and due to \cref{ass:aibi} it holds that $\frac{\ka{0} - \gamma_0}{b_0} + \frac{\pi^2}{4} > 0$
which allows completing the square to get
    \begin{align}
        \dot{W}(t) &\le\left(\lM -\thetam\frac{\pi^2}{4} \right)\ltwonorm{z(\cdot, t)}^2 + \frac{\thetam\pi^4 \bar{z}^2(t)}{16 \left(\frac{\ka{0} - \gamma_0}{b_0} + \frac{\pi^2}{4}\right)}. \nonumber
    \end{align}
    Exploiting again \cref{eq:dc_less_norm} results in
    \begin{align}
        \dot{W}(t) &\le -\tilde{\beta} \ltwonorm{z(\cdot, t)}^2 \label{eq:Wdot_final}
    \end{align}
    where $\tilde{\beta} \coloneqq \thetam\frac{\pi^2}{4} -\lM - \thetam \frac{\pi^4}{16 \left(\frac{\ka{0} - \gamma_0}{b_0} + \frac{\pi^2}{4}\right)}.$  Performing similar steps as those from \cref{eq:betapos} to \cref{eq:ineq_k0} one transforms the inequality $\tilde{\beta} > 0$  to
    \begin{align}
        k_0 - \gamma_0 > \frac{\lM \pi^2}{\thetam \pi^2 - 4 \lM} b_0 - a_0 \label{eq:betatildepos}
    \end{align}
    which is implied by \cref{eq:tuning_k0_mono}. \\
    The inequalities \cref{eq:betatildepos,eq:Wdot_final} imply that $W(t)$ is nonincreasing, which means that
	\begin{align}
		0 \le W(t) \le W(0) \quad \forall t \ge 0. \label{eq:Wbnd}
	\end{align}
	From \cref{eq:Wbnd}, and taking \cref{eq:V3,eq:V7} together with
	\begin{align}
		V_3(t) \le W(t) \le W(0), \\
		V_7(t) \le W(t) \le W(0),
	\end{align}
        into account, one can derive uniform upper bounds for $\int_0^1 \theta(x) z_x^2 ~\dd x$ and $|z(x_i, t)|$, given by
	\begin{align}
		\int_0^1 \theta(x) z^2_x(x,t) ~\dd x \le 2 W(0), \label{eq:bndNORMzx} \\
		|z(0,t)| \le \frac{\Phi_i}{k_i + a_i} + \sqrt{\frac{2b_i}{(k_i + a_i) \theta(x_i)} W(0)}. \label{eq:bndabsZ0}
	\end{align}
	
	\subsection*{Asymptotic state stability}
	In the last step of the proof Barbalat's Lemma is employed in order to show global asymptotic stability in the $L_2$-sense. From \cref{eq:Vdotle0,eq:V1} it is obtained that
	\begin{align}
		\dot{V}(t) &\le -\beta \ltwonorm{z(\cdot, t)}^2 \le -2\beta V_1(t). \label{eq:VdotBarb03}
	\end{align}
	Integrating both sides of \cref{eq:VdotBarb03} from $t=0$ to infinity yields
	\begin{align}
		\lim_{t\to\infty}\{V(0) - V(t)\} \ge 2\beta \int_0^\infty V_1(\tau) ~\dd \tau.\label{eq:VdotBarb04}
	\end{align}
	By virtue of \cref{eq:Vbnd} it follows that
	\begin{align}
		\lim_{t\to\infty}\{V(0) - V(t)\} \ \in \ [0, V(0)]
	\end{align}
	which, considered together with \cref{eq:VdotBarb04}, shows that the integral term in the right-hand side of \cref{eq:VdotBarb04} exists. To apply Barbalat's Lemma to derive that $V_1(t)$ asymptotically converges to zero, it remains to show that $V_1(t)$ is uniformly continuous. The differentiability of $V_1(t)$ and uniform boundedness of $\dot{V}_1(t)$ are sufficient for this. The expression of $\dot{V}_1(t)$ was previously derived in \cref{eq:V1dot}.
	Due to \cref{ass:theta_lambda} it holds that
	\begin{align}
		\lm \ltwonorm{z(\cdot, t)}^2 \le \int_0^1 \lambda(x) z^2(x, t) ~\dd x \le \lambda_\text{M} \ltwonorm{z(\cdot, t)}^2.
	\end{align}
	In view of this and by virtue of \cref{ass:theta_lambda,ass:aibi}, \cref{eq:bndM,eq:bndNORMzx,eq:bndabsZ0,eq:bndNormz} it can be concluded that all terms appearing in the right-hand side of  \cref{eq:V1dot} are uniformly bounded, and so  $\dot{V}_1(t)$  turns out to be uniformly bounded as well. This implies, according to Barbalat's Lemma, that $\lim_{t\to\infty} V_1(t) = \lim_{t\to\infty} \frac{1}{2}\ltwonorm{z(\cdot, t)}^2 = 0$
which concludes the proof.	\hfill \hphantom{} \qed
\end{pf}

\begin{remark}
    The bound $\mathcal{B}$ in \cref{eq:boundB} increases with larger values of $\Phi_i$, which is intuitive. Moreover, for large values of $\alpha_i$, that is, $\min\{\alpha_0, \alpha_1\} > \beta_\mathrm{min}$ (see \cref{eq:rhodef}), the bound $\mathcal{B}$ decreases when $\alpha_i$ decrease. This is also intuitive, as the monodirectional case is approached where essentially $\mathcal{B}\to 0$. However, since the expression for $\mathcal{B}$ contains the parameters $\alpha_i$ in both the numerator and the denominator, the asymptotic convergence result \labelcref{item:monodir} is not obtainable from the stability proof of \labelcref{item:bidir} by letting $\alpha_i \to 0$. \\
    This result shows that smaller adaptation gains $\alpha_i$ generally yield lower ultimate bounds $\mathcal{B}$, though asymptotic convergence is only ensured in the monodirectional adaptation case.
\end{remark}

\section{Experimental validation} \label{sec:application}
\begin{figure}
	\centering
	\footnotesize
	% \includesvg[width=.7\linewidth]{lab_setup_2.svg}
	\def\svgwidth{.7\linewidth}
	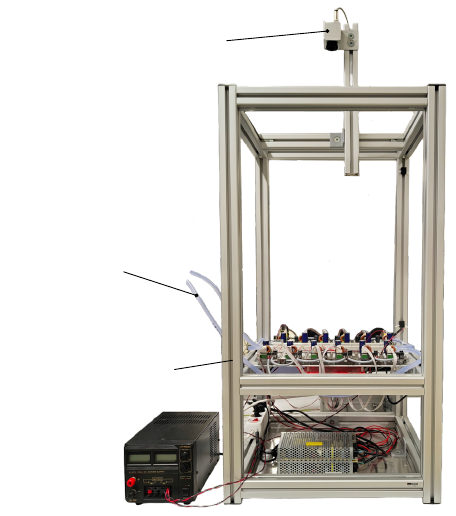
	\caption{A thermal imaging camera measures the temperature of the aluminium beam. The bottom sides of the TEMs are kept at ambient temperature by a water cooling system.}
	\label{fig:lab_setup}
\end{figure}

\begin{figure}
	\footnotesize
	\centering
	% \includesvg[width=.8\columnwidth]{boundary_control_dist_highlight_hint.svg}
	\def\svgwidth{.8\columnwidth}
	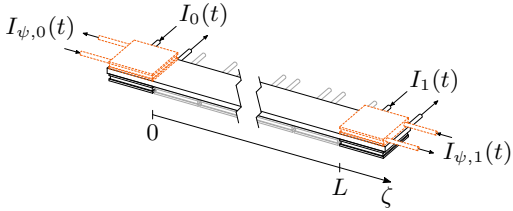
	\caption{The aluminium beam is cooled/heated by the outermost TEMs, used as boundary actuators, whereas the inner TEMs are inactive during all conducted experiments. The TEMs highlighted by orange dashed lines are used for imposing matched boundary disturbances.}
	\label{fig:boundary_control}
\end{figure}

The applicability of the presented control law is demonstrated on the laboratory setup depicted in \cref{fig:lab_setup}. The core of the setup consists of an aluminium beam which is mounted on an array of TEMs that allow thermal actuation of the beam. When applying a voltage to the terminals of a TEM, the resulting electrical current induces a temperature difference between the two sides of the TEM due to the Peltier effect. 
Since the Peltier effect depends on the electrical current, dedicated controllers are integrated into the laboratory setup which regulate the currents flowing across the TEMs to the desired values. The dynamics of these current loops are considered fast in comparison to the thermal dynamics of the process, hence, they are neglected in the control design. The setpoints for the current controllers are, therefore, considered the manipulable control inputs of the plant under consideration. \\
The outermost TEMs are used as actuators, as shown in \cref{fig:boundary_control}, whereas the TEMs mounted inside the domain $\zeta \in [0, L]$ are never activated. Two additional TEMs are located at the boundaries on top of the beam, whose purpose is to introduce specifiable boundary disturbances. \\
The temperature of the beam is measured by a thermal imaging camera, mounted at the top of the setup and observing the beam from a bird's eye perspective. This allows temperature measurement over the whole domain which is beneficial for analysis purposes, however, the controller uses measurements at the boundaries $\zeta = 0$ and $\zeta = L$ only.

\subsection{Modelling}
    This section derives a mathematical model of the thermal system under consideration. It begins with the formulation of heat conduction within a metallic beam and dissipation to the environment, followed by a detailed model of the thermoelectric actuators.
	\subsubsection{Heat conduction and dissipation}
	\begin{table}
		\centering
		\caption{Parameters of the aluminium beam.} \label{tab:beam_params}
		\vskip\baselineskip
		\begin{tabular}{@{}clrc@{}}
			\toprule
			\multicolumn{1}{l}{Parameter} & Description            & \multicolumn{1}{l}{Value} & \multicolumn{1}{l}{Unit}       \\ \midrule
			$\Lal$                        & Length                 & 315                       & mm                             \\
			$\hal$                        & Thickness              & 3                         & mm                             \\
			$\bal$                        & Width                  & 25                        & mm                             \\
			$k$                           & Thermal conductivity   & 209                       & $\frac{\text{W}}{\text{K}\cdot\text{m}}$   \\
			$\rho$                        & Density                & 2700                      & $\frac{\text{kg}}{\text{m}^3}$ \\
			$\cp$                         & Specific heat capacity & 898                       & $\frac{\text{J}}{\text{kg}\cdot\text{K}}$  \\ \bottomrule
		\end{tabular}
	\end{table}
	The considered beam is made of an aluminium alloy \texttt{EN AW-6060}. Its temperature profile $T(\zeta,t)$ is governed by the heat PDE
	\begin{align}
		\rho \cp T_t(\zeta,t) &= k T_{\zeta\zeta}(\zeta,t) - h \frac{A}{V} \left[T(\zeta,t) - \Ta \right] \label{eq:model_PDE}
	\end{align}
	with the spatial coordinate $\zeta \in (0,L)$ and constant material parameters as described in \cref{tab:beam_params}. This is a special case of \cref{eq:PDE} with constant, rather than spatially-varying, material parameters. \\
    As illustrated in \cref{fig:boundary_control}, the outermost TEMs are used as boundary actuators, hence their lengths $\ltem$ are subtracted from the beam length $\Lal$ to get the effective length $L = \Lal - 2 \ltem$
of the spatial domain.
	The second term in the right-hand side of \cref{eq:model_PDE} models the convective heat losses to the environment with heat transfer coefficient $h = 69.3\,\frac{\text{W}}{\text{K}\cdot\text{m}^2}$, where $\Ta$ is the ambient temperature, and with the surface to volume ratio $\frac{A}{V} = \frac{2 L \bal + 2 L \hal}{L \bal \hal}$.
With the definitions
	\begin{align}
		\thetab &\coloneqq \frac{k}{\rho \cp}, \quad	\lambda \coloneqq -\frac{h}{\rho \cp} \frac{A}{V} \label{eq:lambda_def}
	\end{align}
	the PDE \cref{eq:model_PDE} is rewritten as
	\begin{align}
			T_t(\zeta,t) &= \thetab T_{\zeta\zeta}(\zeta,t) + \lambda \left[T(\zeta,t) - \Ta \right]. \label{eq:PDEmodel}
	\end{align}
	Heat can enter the domain at the boundaries, which is described by the BCs
	\begin{subequations}
	\begin{align}
		T_\zeta(0,t) &= -\frac{1}{k} \left[\dot{q}_0(t)  + \qpsi{0}(t) \right], \label{eq:T_BC0} \\
		T_\zeta(L, t) &= \frac{1}{k} \left[\dot{q}_1(t) + \qpsi{1}(t) \right] \label{eq:T_BCL}
	\end{align}
	\end{subequations}
	where $\dot{q}_i(t)$ and $\dot{q}_{\psi,i}(t)$ are imposed and unknown heat flux densities, respectively. The unknown heat flux densities $\dot{q}_{\psi,i}(t)$ arise from model uncertainties, but later they will also be generated by the additionally mounted TEMs on top of the beam to test the robustness of the closed-loop system.

		\subsubsection{Actuator model}
	The boundary-located TEMs generate the heat fluxes $\Qtem{i}(t)$ depending on the currents $I_i(t)$ that are supplied to them.
	There are three main effects which describe the behavior of the TEM, namely heat conduction, the Peltier effect and Joule losses \cite{Lineykin2007}. These are summarized in the relation
	\begin{align}
		\Qtem{i} &= \frac{\Ta - T(\zeta_i,t)}{\Rth{i}} + \sigma_i I_i(t) T(\zeta_i,t) + \frac{R_i}{2} I_i^2(t), \label{eq:TEMmodel}
	\end{align}
	 where $i \in \{0, 1\}$, $\zeta_0=0$, $\zeta_1=L$, and with the parameters given in \cref{tab:tem_parameters}.
	\begin{table}
		\centering
		\caption{Parameters of the TEMs.} \label{tab:tem_parameters}
		\vskip\baselineskip
		\begin{tabular}{@{}clrrc@{}}
			\toprule
			\multicolumn{1}{l}{}          &                       & \multicolumn{2}{c}{Value}                             & \multicolumn{1}{l}{}         \\ \midrule
			\multicolumn{1}{l}{Parameter} & Description           & \multicolumn{1}{c}{$i=0$} & \multicolumn{1}{c}{$i=1$} & \multicolumn{1}{l}{Unit}     \\ \midrule
			$\Rth{i}$                  & Thermal resistance    & 2.21                      & 2.15                      & $\frac{\text{K}}{\text{W}}$  \\
			$\sigma_i$                    & Seebeck parameter     & 0.037                     & 0.044                     & $\frac{\text{mV}}{\text{K}}$ \\
			$R_i$                         & Electrical resistance & 10.2                      & 13.5                      & $\Omega$                     \\ 
			$\ltem$                       & Length				  & 25                        & 25                        & mm						     \\ \bottomrule
		\end{tabular}
	\end{table}
	The heat fluxes $\Qtem{i}$ split up into the parts $\dot{Q}_i$ which enter the domain and the parts $\Qloss{i}(t)$ which leave to the environment, i.e.
	\begin{align}
		\Qtem{i}(t) &= \dot{Q}_i(t) + \Qloss{i}(t) \label{eq:Qkirchhoff}
	\end{align}
	Rearranging \cref{eq:Qkirchhoff} and modelling the losses via $\Qloss{i}(t) = h \Atemloss \left[T(\zeta_i,t) - \Ta\right]$
where $\Atemloss = \ltem \bal + 2 \ltem \hal + \bal \hal$
leads to
	\begin{align}
		\dot{Q}_i(t) &= \Qtem{i}(t) - h \Atemloss \left[T(\zeta_i,t) - \Ta\right]. \label{eq:Qdot}
	\end{align}
	Dividing by the cross-sectional area $\Acs = \bal \hal$	gives the heat flux densities
	\begin{align}
		\dot{q}_i(t) &= \frac{\dot{Q}_i(t)}{\Acs}, \quad i \in \{0, 1\}. \label{eq:heatfluxdensity}
	\end{align}
	Inserting \cref{eq:TEMmodel} into \cref{eq:Qdot} and considering \cref{eq:heatfluxdensity} and $\Qpsi{i}(t) \coloneqq \qpsi{i}(t) \Acs$ for $i \in \{0, 1\}$
together with \cref{eq:T_BC0,eq:T_BCL} leads to the Robin BCs
\begin{subequations}\label{eq:BCsmodel}
	\begin{align} 
		-\bb_0 T_\zeta(0,t) + a_0 \left[T(0,t) - \Ta\right] &= I_0(t) T(0,t) \alignbreak \hspace{-15mm} + \frac{1}{2} \frac{R_i}{\sigma_0} I_0^2(t) + \frac{\Qpsi{0}(t)}{\sigma_0}, \label{eq:BC0model} \\
		\bb_1 T_\zeta(1,t) + a_1 \left[T(L,t) - \Ta\right] &= I_1(t) T(L,t) \alignbreak \hspace{-15mm} + \frac{1}{2}\frac{R_i}{\sigma_1} I_1^2(t) + \frac{\Qpsi{1}(t)}{\sigma_1} \label{eq:BCLmodel}
	\end{align}
	\end{subequations}
	where
	\begin{align}
		a_i \coloneqq \frac{1}{\sigma_i}\left(\frac{1}{\Rth{i}} + h \Atemloss\right), \quad
		\bb_i \coloneqq \frac{k \Acs}{\sigma_i}. \label{eq:ai}
	\end{align}
	The overall model is therefore given by the PDE \cref{eq:PDEmodel} together with Robin BCs \cref{eq:BC0model,eq:BCLmodel}.

    \subsection{Model identification and validation}
\begin{figure}
		\centering
		\begin{subfigure}[b]{0.48\textwidth}
			\centering
			\includegraphics{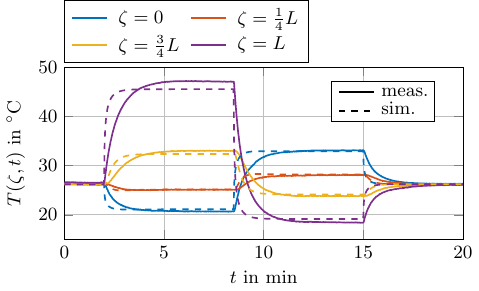}
			\caption{The measurements are compared to the simulation results at different locations.}
			\label{fig:validation_exp}
		\end{subfigure}
		\hfill
		\begin{subfigure}[b]{0.48\textwidth}
			\centering
			\includegraphics{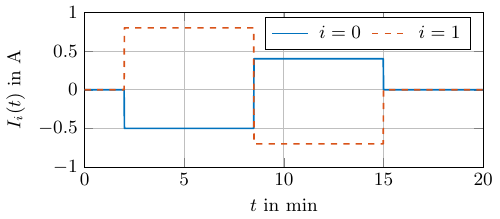}
			\caption{A step-wise current is applied to the setup during the validation experiment.}
			\label{fig:validation_exp_input}
		\end{subfigure}
		\caption{Validation of the model by comparing an experiment with a corresponding simulation.}
		\label{fig:validation_exp_combined}
	\end{figure}

	The parameters of the TEMs in \cref{tab:tem_parameters} are obtained by performing an identification experiment and applying an optimization strategy. A similar approach yields the estimated heat transfer coefficient value
	\begin{align}
		h &= 69.3\,\frac{\text{W}}{\text{K}\cdot\text{m}^2}.
	\end{align}
A different excitation dataset is used to validate the model including the identified parameters. 
    The measurements of this validation experiment with a step-wise input are depicted in \cref{fig:validation_exp} along with simulation results. The applied current is given in \cref{fig:validation_exp_input}. The comparison shows that the model resembles the real-world setup reasonably well, especially in steady state. In the transient phases the simulation behaves quicker than the physical setup which is suspected to originate from the unmodelled heat conduction process occurring at each boundary.

\subsection{Feedback loop design}
	The goal is to achieve a specific temperature profile in the presence of external matched disturbances, i.e., an equilibrium of \cref{eq:PDEmodel,eq:BCsmodel} should be robustly stabilized. An input linearization is designed in \cref{sec:compensation} which deals with the nonlinearities in \cref{eq:BCsmodel}. Based on specified boundary temperatures, a corresponding equilibrium is calculated in \cref{sec:equi}, which is followed by the derivation the corresponding error dynamics in \cref{sec:errordyn}.

	\subsubsection{Input linearization} \label{sec:compensation}
\begin{figure}
		\centering
\includegraphics{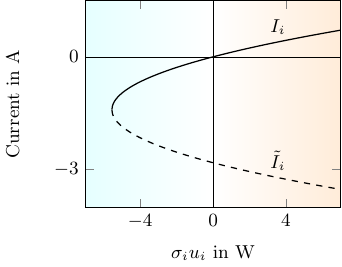}
		\caption{There are two possibilities for the current to achieve a certain heat flux $\sigma_i u_i$. The current is plotted over $\sigma_i u_i$ instead of $u_i$ for better physical interpretation.}
		\label{fig:compensation_sketch}
	\end{figure}

	New inputs
	\begin{align}
		u_i(t) &= I_i(t) T(\zeta_i,t) + \frac{1}{2} \frac{R}{\sigma_i} I_i^2(t) \label{eq:newinput}
	\end{align}
	and disturbances $\psi_i(t) = \frac{\Qpsi{i}(t)}{\sigma_i}$
for $i \in \{0, 1\}$ are defined, which renders the BCs \cref{eq:BC0model,eq:BCLmodel} to
	\begin{subequations}
	\begin{align}
		-\bb_0 T_\zeta(0,t) + a_0 T(0,t) &= u_0(t) + \psi_0(t) + a_0\Ta, \label{eq:BC0_1}\\
		\bb_1 T_\zeta(L,t) + a_1 T(L,t) &= u_1(t) + \psi_1(t) + a_1\Ta. \label{eq:BCL_1}
	\end{align}
	\end{subequations}
        By inverting \cref{eq:newinput} the compensation function
	\begin{align}
		I_i(t) &= -\frac{\sigma_i}{R_i}\left(T(\zeta_i, t) - \sqrt{T^2(\zeta_i, t)+2\frac{R_i}{\sigma_i}u_i(t)}\right) \label{eq:compensation}
	\end{align}
     is obtained. Note that \cref{eq:newinput} is a quadratic function, hence, there are in fact two solutions for $I_i(t)$. The solution \cref{eq:compensation} is sketched in \cref{fig:compensation_sketch} along with the other possible solution $\tilde{I}_i(t)$ for a reasonable choice of parameters. Physical intuition can be obtained for $\tilde{I}_i(t)$ when considering that with high negative currents the (always non-negative) Joule losses dominate the Peltier effect. This way even a net heating can be achieved with a negative current. The operating mode described by $\tilde{I}_i(t)$, however, is not favorable as the current applied to the TEMs would be high in magnitude permanently. Thus, the solution \cref{eq:compensation} is chosen as the compensation function.

	\subsubsection{Reference equilibrium} \label{sec:equi}
		The equilibrium to be stabilized is calculated with the assumption $\psi_0(t) = \psi_1(t) = 0$.
\begin{subequations} \label{eq:equi_sys}
		Setting the time derivative in \cref{eq:PDEmodel} to zero results in the boundary value problem
			\begin{align}
				0 = \thetab \Te''(\zeta) + \lambda &\left[\Te(\zeta) - \Ta \right], \label{eq:equi_ode} \\
				-\bb_0 \Te'(0) + a_0 \Te(0) &= \une + a_0\Ta, \label{eq:equi_BC0} \\
				\bb_1 \Te'(L) + a_1 \Te(L) &= \uoe + a_1\Ta \label{eq:equi_BCL}
			\end{align}
		\end{subequations}
		where $\Te'$ and $\Te''$ denote the first and second spatial derivative of $\Te$, respectively.
		In order to solve the boundary value problem, the transformation
		\begin{align}
			\ze(\zeta) \coloneqq \Te(\zeta) - \Ta \label{eq:equistate}
		\end{align}
		is performed, which yields
		\begin{subequations}
			\begin{align}
				0 &= \thetab \ze''(\zeta) + \lambda \ze(\zeta), \label{eq:trans_equi_ode} \\
				-\bb_0\ze'(0) + a_0 \ze(0) &= \une, \label{eq:BC0e} \\
				\bb_1\ze'(L) + a_1 \ze(L) &= \uoe. \label{eq:BC1e}
			\end{align}
		\end{subequations}
		The general solution of \cref{eq:trans_equi_ode} for  $\lambda < 0$ is (see \cite{Zaitsev2002})
		\begin{align}
			\ze(\zeta) &= C_1 \sinh\left(\omega \zeta\right) + C_2 \cosh\left(\omega \zeta\right) \label{eq:general_sol}
		\end{align}
        with $\omega \coloneqq \ltfac$.
		Considering \cref{eq:BC0e,eq:BC1e} together with the derivative of \cref{eq:general_sol}
results in the linear system of equations
		\begin{align}
			\begin{bmatrix}
				-\bb_0\omega & a_0\\ \alpha_{21} & \alpha_{22}
			\end{bmatrix} \begin{bmatrix}
				C_1 \\ C_2
			\end{bmatrix} &= \begin{bmatrix}
				\une \\ \uoe
			\end{bmatrix} \label{eq:Csys}
		\end{align}
		where
		\begin{align}
			\alpha_{21} &= a_1\sinh\left(L\omega\right)+\bb_1\omega\cosh\left(L\omega\right), \\
			\alpha_{22} &= a_1\cosh\left(L\omega\right)+\bb_1\omega\sinh\left(L\omega\right).
		\end{align}
        The solution of \cref{eq:Csys} is given by
\begin{subequations}
			\begin{align}
				C_1 &=-\frac{a_1\une\cosh\left(L\omega\right)-a_0\uoe+\bb_1\une\omega\sinh\left(L\omega\right)}{d_A}, \\
				C_2 &= \frac{\bb_0\uoe\omega+a_1\une\sinh\left(L\omega\right)+\bb_1\une\omega\cosh\left(L\omega\right)}{d_A}
			\end{align}
		\end{subequations}
		where
		\begin{align} d_A &= (a_0a_1+\bb_0\bb_1\omega^2)\sinh\left(L\omega\right)\alignbreak+\left(a_0\bb_1 + a_1 \bb_0\right)\omega\cosh\left(L\omega\right).
		\end{align}
		By considering \cref{eq:equistate,eq:general_sol} the equilibrium of \cref{eq:PDEmodel,eq:BC0_1,eq:BCL_1} is given by
		\begin{align}
			\Te(\zeta) &= C_1 \sinh\left(\omega \zeta\right) + C_2 \cosh\left(\omega \zeta\right) + \Ta.
		\end{align}
		Given the two desired boundary temperatures $\Te(0) \eqqcolon \Ten$ and ${\Te(L) \eqqcolon \TeL}$
the inputs of the equilibrium are determined by
		\begin{align}
			\une(\Ten, \TeL) &= \frac{1}{\sinh\left(L\omega\right)} \bigg( (\Ten - \Ta) [a_0\sinh\left(L\omega\right) \alignbreak + \bb_0\omega\cosh\left(L\omega\right)] -\bb_0\omega(\TeL - \Ta) \bigg), \nonumber \\
			\uoe(\Ten, \TeL) &= \frac{1}{\sinh\left(L\omega\right)} \bigg( (\TeL - \Ta) [a_1\sinh\left(L\omega\right) \alignbreak + \bb_1\omega\cosh\left(L\omega\right) ]-\bb_1\omega(\Ten - \Ta) \bigg). \nonumber
\end{align}

	\subsubsection{Error dynamics} \label{sec:errordyn}
\begin{figure}
		\scriptsize
		\centering
		% \includesvg[width=.95\columnwidth]{control_loop.svg}
		\def\svgwidth{.95\columnwidth}
		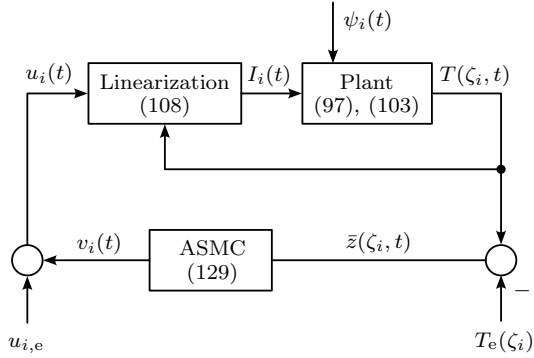
		\caption{The feedback loop stabilizes the equilibrium $(\Te, u_{i,\text{e}})$ and is robust against external, matched disturbances $\psi_i(t)$, $i \in \{0,1\}$.} \label{fig:feedback_loop}
	\end{figure}

	The goal is to drive the difference between the state $T(\zeta, t)$ and the desired equilibrium $\Te(\zeta)$ to zero, hence, the error variable
	\begin{align}
		\errvar(\zeta,t) &\coloneqq T(\zeta,t) - \Te(\zeta)
	\end{align}
	is defined.
	Performing the state transformation by considering
	\begin{subequations}
	\begin{align}
		\errvar_t(\zeta,t) &= T_t(\zeta,t), \label{eq:errt} \\
		\errvar_\zeta(\zeta,t) &= T_\zeta(\zeta,t) - \Te'(\zeta), \label{eq:erry} \\
		\errvar_{\zeta\zeta}(\zeta,t) &= T_{\zeta\zeta}(\zeta,t) - \Te''(\zeta) \label{eq:erryy}
	\end{align}
	\end{subequations}
	and inserting \cref{eq:errt,eq:erry,eq:erryy} into \cref{eq:PDEmodel} yields
	\begin{align}
		\errvar_t(\zeta,t) &= \thetab \errvar_{\zeta\zeta}(\zeta,t) + \lambda \errvar(\zeta,t) \alignbreak + \underbrace{\thetab \Te''(\zeta) + \lambda \left[\Te(\zeta) - \Ta\right]}_{= 0}. \label{eq:e_pde_1}
	\end{align}
	Due to \cref{eq:equi_ode} the last two terms in \cref{eq:e_pde_1} are zero.
	Inserting  \cref{eq:errt,eq:erry} into \cref{eq:BC0_1} and considering \cref{eq:equi_BC0} gives
	\begin{align}
-\bb_0 \errvar_\zeta(0,t) + a_0 \errvar(0,t) &= u_0(t) -\une + \psi_0(t).
	\end{align}
	Similarly, by virtue of \cref{eq:errt,eq:erry,eq:BCL_1,eq:equi_BCL},
	\begin{align}
		\bb_1 \errvar_\zeta(L,t) + a_1 \errvar(L,t) &= u_1(t) - \uoe + \psi_1(t)
	\end{align}
	is obtained for the BC at $\zeta = L$. The error dynamics can therefore be summarized by
		\begin{subequations}
		\begin{align}
			\errvar_t(\zeta,t) &= \thetab \errvar_{\zeta\zeta}(\zeta,t) + \lambda \errvar(\zeta,t), \quad \zeta \in (0, L), \label{eq:pde_err}\\
			-\bb_0 &\errvar_\zeta(0,t) + a_0 \errvar(0,t) = v_0(t) + \psi_0(t), \label{eq:bc0_err} \\
			\bb_1 &\errvar_\zeta(L,t) + a_1 \errvar(L,t) = v_1(t) + \psi_1(t) \label{eq:bcL_err}
		\end{align}
	\end{subequations}
	where
	\begin{align}
		v_0(t) &\coloneqq u_0(t) - \une, &
		v_1(t) &\coloneqq u_1(t) - \uoe.
	\end{align}
	Next, the coordinate transformation $x = \frac{\zeta}{L}$ is performed in order to bring the spatial domain of \cref{eq:pde_err,eq:bc0_err,eq:bcL_err} from $\zeta \in [0,L]$ to $x \in [0,1]$. From
	\begin{align}
		z(x,t) &= \errvar(\zeta(x), t) = \errvar(xL, t) \label{eq:z_zbar}
	\end{align}
	it follows that
	\begin{subequations}
	\begin{align}
		z_x(x,t) &= \errvar_\zeta(\zeta,t) \dv{\zeta}{x} = \errvar_\zeta(\zeta,t) L \\
		z_{xx}(x,t) &= \errvar_{\zeta\zeta}(\zeta,t) L^2 \\
		z_t(x,t) &= \errvar_t(\zeta,t)
	\end{align}
	\end{subequations}
	and so the PDE \cref{eq:pde_err} becomes
	\begin{subequations}
	\begin{align}
		z_t(x,t) &= \frac{\thetab}{L^2} z_{xx}(x,t) + \lambda z(x,t), \quad x \in (0,1). \label{eq:pde_final}
	\end{align}
	The BCs \cref{eq:bc0_err,eq:bcL_err} are rendered to
	\begin{align}
		-\frac{\bb_0}{L} z_x(0, t)  + a_0 z(0,t) &= v_0(t) + \psi_0(t), \label{eq:bc0_final} \\
		\frac{\bb_1}{L} z_x(1,t) + a_1 z(1,t) &= v_1(t) + \psi_1(t). \label{eq:bc1_final}
	\end{align}
	\end{subequations}
	With $\theta = \frac{\thetab}{L^2}$	and $b_i = \frac{\bb_i}{L}$ for $i \in \{0,1\}$
the boundary value problem \cref{eq:pde_final,eq:bc0_final,eq:bc1_final} is equivalent to the one in \cref{sec:problem_formulation} for which 
	the stability proof is conducted, hence, the adaptive sliding mode control (ASMC) law from \cref{sec:controller} can be applied.
	By considering \cref{eq:z_zbar} the control law \cref{eq:control} and adaptation law \cref{eq:control_adapt} in terms of the error coordinates read as
        \begin{subequations} \label{eq:control_error_full}
    	\begin{align}
    		v_i(t) &= -\propgain_i \errvar(\zeta_i,t) - M_i(t) \sign{\errvar(\zeta_i,t)},  \label{eq:control_error} \\
            \dot{M}_i(t) &= -\alpha_i M_i(t) + \gamma_i |\errvar(\zeta_i,t)| \label{eq:control_adapt_error}
    	\end{align}
        \end{subequations}
	which results in the overall feedback loop depicted in \cref{fig:feedback_loop}.
	
\subsection{Closed-loop experiments}
\begin{figure}
	\centering
	\begin{subfigure}[b]{0.48\textwidth}
		\centering
		\includegraphics{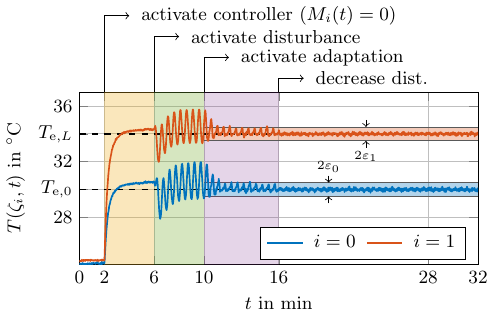}
		\caption{Monodirectional adaptation ($\alpha_0 = \alpha_1 = 0$)}
	\end{subfigure}
	\phantom{a}
	\begin{subfigure}[b]{0.48\textwidth}
		\centering
		\includegraphics{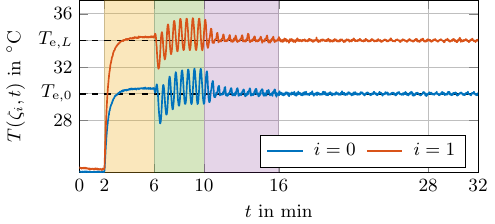}
		\caption{Bidirectional adaptation ($\alpha_0 > 0$, $\alpha_1 > 0$)}
	\end{subfigure}
    \caption{The experiment is structured into multiple phases to better highlight the robustification through the discontinuous switching action. It is clearly visible how the oscillations from the matched disturbances are attenuated.}
    \label{fig:cl_sliding_vars}
\end{figure}

\begin{figure}
	\centering
	\includegraphics{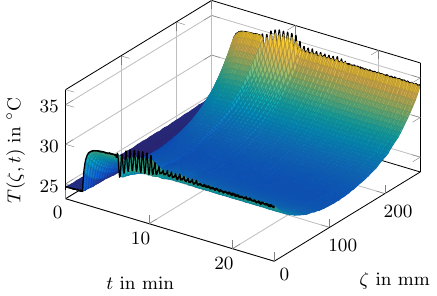}
	\caption{Spatiotemporal temperature profile during the experiment. The reference equilibrium is stabilized despite the presence of matched disturbances imposed by the additionally mounted TEMs.}
	\label{fig:cl_surface}
\end{figure}

\begin{figure}
	\centering
	\includegraphics{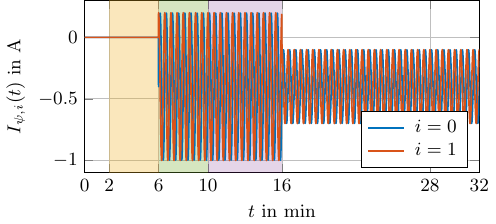}
	\caption{Matched disturbances imposed by means of the additionally mounted TEMs.}
	\label{fig:cl_current_psi}
\end{figure}

\begin{figure}
	\centering
	\begin{subfigure}[b]{0.48\textwidth}
		\centering
		\includegraphics{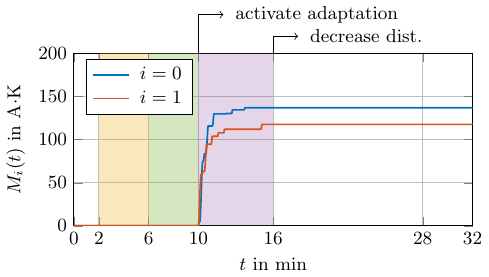}
		\caption{Monodirectional adaptation ($\alpha_0 = \alpha_1 = 0$)}
	\end{subfigure}
	\phantom{a}
	\begin{subfigure}[b]{0.48\textwidth}
		\centering
		\includegraphics{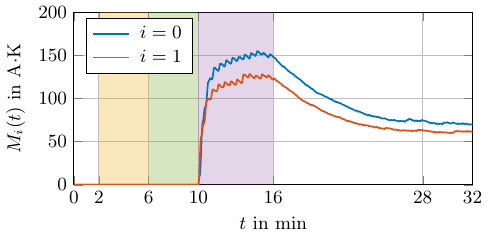}
		\caption{Bidirectional adaptation ($\alpha_0 > 0$, $\alpha_1 > 0$)}
	\end{subfigure}
    \caption{The adaptation is enabled at $t = 10\,\text{min}$. At $t = 16\,\text{min}$ the amplitude of the disturbance decreases which results in decreasing adaptive gains in the bidirectional adaptation case. In the monodirectional adaptation case, however, the adaptive gains maintain their previously attained values.}
	\label{fig:cl_M}
\end{figure}

\begin{figure}
	\centering
	\begin{subfigure}[b]{0.48\textwidth}
		\centering
		\includegraphics{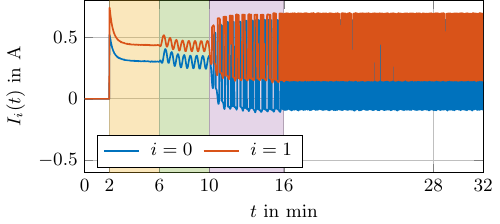}
		\caption{Monodirectional adaptation ($\alpha_0 = \alpha_1 = 0$)}
	\end{subfigure}
	\phantom{a}
	\begin{subfigure}[b]{0.48\textwidth}
		\centering
		\includegraphics{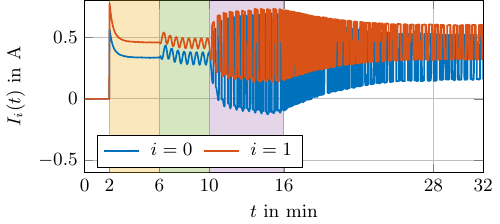}
		\caption{Bidirectional adaptation ($\alpha_0 > 0$, $\alpha_1 > 0$)}
	\end{subfigure}
    \caption{Manipulable input currents of the actuator TEMs.}
    \label{fig:cl_current}
\end{figure}

\begin{figure}
	\centering
	\includegraphics{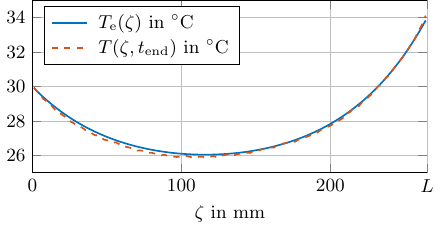}
	\caption{The measured temperature profile $T(\zeta, t_\text{end})$ at the end of the experiment closely matches the imposed reference equilibrium $\Te(\zeta)$. }
	\label{fig:cl_equi}
\end{figure}

Closed-loop experiments are performed on the laboratory setup in order to demonstrate the applicability of the proposed control schemes and to compare the mono- and bidirectional algorithm. In such a real-world application imperfections may occur which highlight a major shortcoming of the monodirectional adaptation algorithm. These imperfections can be measurement noise, but also chattering which originates from unmodelled dynamics or the discrete-time implementation of the controller. Consequently, the convergence of the boundary errors $\errvar(\zeta_i, t)$ towards zero is no longer achieved. With the adaptation law \cref{eq:control_adapt} with $\alpha_0=\alpha_1=0$ this would cause an unbounded drift of the switching gains $M_i(t)$. To circumvent this effect, for the experiment the adaptation law \cref{eq:control_adapt} is subject to a dead-zone redesign according to
\begin{align}
	\dot{M}_i(t) &= \begin{cases}
		\gamma_i |\errvar(\zeta_i, t)| \; &\text{if }  |\errvar(\zeta_i, t)| > \varepsilon_i \\
		0 & \text{else}
	\end{cases}, \label{eq:control_adapt_error_extended}
\end{align}
$i\in\{0,1\}$, for the monodirectional case. That is, the adaptation is paused when the boundary error lies within a band around zero of width $\varepsilon_i$. With the bidirectional adaptation algorithm such a modification is not necessary, as boundedness of the adaptive gains $M_i(t)$ is always guaranteed for bounded error variables $\errvar(\zeta_i, t)$. \\ 
To better demonstrate the robustness of the control scheme, the additionally mounted TEMs on top of the beam are used to impose heat fluxes $\Qpsi{i}(t)$ acting as matched disturbances. The currents applied to these TEMs are denoted as $I_{\psi,i}(t)$, $i \in \{0,1\}$. \\
Two experiments were conducted which differ only in the used adaptation algorithm.
The desired boundary temperatures are ${\Ten = 30\,^\circ\text{C}}$ and ${\TeL = 34\,^\circ\text{C}}$ and the ambient temperature is measured to be ${\Ta \approx 24.6\,^\circ\text{C}}$ for the experiment with the monodirectional adaptation algorithm and ${\Ta \approx 24.1\,^\circ\text{C}}$ for the experiment with the bidirectional adaptation algorithm. \\
The parameters specifying the integration deadbands for the monodirectional adaptation algorithm are chosen as $\varepsilon_0 = \varepsilon_1 = 0.5\,^\circ\text{C}$. The sample time of the controller is $T_\text{s} = 0.1\,\text{s}$. The adaptation gains are chosen as
\begin{align}
	\gamma_i &= \begin{cases}
		0\,\frac{\text{A}}{\text{s}} \quad &\text{for } 0 \le t < 10\,\text{min} \\
		4\,\frac{\text{A}}{\text{s}} &\text{for } t \ge 10\,\text{min}
	\end{cases}, \label{eq:gamma_exp}
\end{align}
i.e., there is no adaptation during $0 \le t < 10\,\text{min}$. For the bidirectional adaptation case the parameters $\alpha_i$ are chosen as $\alpha_0 = \alpha_1 = \frac{1}{300}\,\frac{1}{\text{s}}$. The initial values of the adaptive gains are chosen in accordance with \cref{eq:tuning_M0} as $M_{0,0} = M_{0,1} = 0$.
For the setup at hand the reaction term of \cref{eq:pde_final} is stabilizing since $\lambda < 0$ holds, as it can be seen in \cref{eq:lambda_def}. Furthermore, the parameters $a_0$ and $a_1$ corresponding to the Dirichlet part of the Robin BCs \cref{eq:bc0_final,eq:bc1_final} are positive (see \cref{eq:ai}). Hence, a choice of $k_i = 15\,\text{A}$ satisfies the parameter conditions \cref{eq:tuning_k_mono,eq:tuning_k_bi}. \\
The step-wise choice of \cref{eq:gamma_exp} is made to structure the conducted experiment into multiple phases, which highlights the robustness gained from the discontinuous switching component in \cref{eq:control}. These phases are illustrated in \cref{fig:cl_sliding_vars}, showing the evolution of the boundary temperatures for both experiments. The spatiotemporal evolution of the temperature $T(\zeta, t)$ during the first minutes of the experiment with monodirectional adaptation is visualized in \cref{fig:cl_surface}. \\
At $t =  2\,\text{min}$ the control law \cref{eq:control_error} is activated while the adaptation \cref{eq:control_adapt_error} respectively \cref{eq:control_adapt_error_extended} is kept deactivated according to \cref{eq:gamma_exp}. Since the adaptive gains $M_i(t)$ are initialized with zero, proportional feedback is applied only. Starting from ambient temperature, the system heats up toward the target values $\Ten$ and $\TeL$, however, the desired temperatures are not attained exactly. \\
At $t =  6\,\text{min}$ disturbances are applied through the additionally mounted TEMs. The chosen disturbance currents are visualized in \cref{fig:cl_current_psi}: they are identically zero for $t < 6\,\text{min}$, they are sinusoidal signals for $6\,\text{min} \le t < 16\,\text{min}$, and they are still sinusoidal, but with reduced amplitude, for $ t \geq 16 \text{min}$. The influence of these disturbances is clearly visible in \cref{fig:cl_sliding_vars}, manifesting as oscillations in the temperatures. \\
At $t = 10\,\text{min}$ the previously deactivated adaptation \cref{eq:control_adapt_error} respectively \cref{eq:control_adapt_error_extended} is engaged, leading to a rise of $M_{i}(t)$ and enabling the discontinuous feedback of \cref{eq:control_error}. The evolution of the adaptive switching gains is depicted in \cref{fig:cl_M}.
The discontinuous control action suppresses the oscillations originating from the imposed disturbances. In the monodirectional case the adaptation stops when the boundary temperatures no longer leave the vicinities around $\Ten$ and $\TeL$ defined by the parameters $\varepsilon_i$. The adaptive switching gains settle to constant values which ensure sufficient disturbance attenuation such that the $\varepsilon_i$ vicinity is maintained. In the bidirectional case the choice of parameters leads to a similar disturbance attenuation. \\
At $t = 16\,\text{min}$ the amplitude of the disturbance is decreased. In the bidirectional case this leads to a decrease of the adaptive gains $M_i(t)$. In the monodirectional case the adaptive gains remain at their previously attained values. \\
In order to compare the residual high-frequency oscillations in the final part of the experiments, the energy-like measure $E = \sum_j (\Ten - T_j)^2$ is calculated from the temperature samples $T_j = T(\zeta_0, jT_\text{s})$ of the time span $28\,\text{min} \le j T_\text{s} \le 32\,\text{min}$ for both the monodirectional and bidirectional adaptation case. This results in $E = 16\,\text{K}^2$ for the monodirectional case and $E = 10.6\,\text{K}^2$ for the bidirectinonal case and therefore highlights a key benefit of the bidirectional adaptation algorithm. \\
\Cref{fig:cl_current} depicts the current of the actuator TEMs and \cref{fig:cl_equi} compares the calculated reference equilibrium to the measured temperature at the last time step, i.e., $t = t_\text{end} = 32\,\text{min}$. The data of the experiment with the monodirectional adaptation algorithm is used for this comparison.

\section{Conclusions} \label{sec:conclusions}
A sliding-mode--based adaptive boundary control law is proposed to stabilize a diffusion process with unknown, spatially varying diffusivity and a potentially destabilizing reaction term with unknown, spatially varying reaction coefficient. Stability properties of the closed-loop system are investigated via Lyapunov-based analyses for two classes of adaptation algorithms of monodirectional and bidirectional type. 
Real-world application of both algorithms in a thermal control setting showed that the seemingly stronger asymptotic stability properties of the monodirectional case are weakened, as practical implementation requires the dead-zone modification of the adaptation law to avoid the otherwise inevitable unbounded drift of the control gain due to effects such as measurement noise. The comparison highlighted the benefit of the bidirectional adaptation with regard to chattering avoidance. An interesting extension would be that of considering unmatched in-domain disturbances to be possibly attenuated by collocated in-domain actuators, as done in the non-adaptive setting of \cite{Pisano2017}, to achieve a closed-loop ISS property.

\bibliographystyle{ieeetr}
\bibliography{references}

\end{document}

%% file: lab_setup_2_svg-tex.pdf_tex
%% Creator: Inkscape 1.3.2 (1:1.3.2+202311252150+091e20ef0f), www.inkscape.org
%% PDF/EPS/PS + LaTeX output extension by Johan Engelen, 2010
%% Accompanies image file 'lab_setup_2_svg-tex.pdf' (pdf, eps, ps)
%%
%% To include the image in your LaTeX document, write
%%   \input{<filename>.pdf_tex}
%%  instead of
%%   \includegraphics{<filename>.pdf}
%% To scale the image, write
%%   \def\svgwidth{<desired width>}
%%   \input{<filename>.pdf_tex}
%%  instead of
%%   \includegraphics[width=<desired width>]{<filename>.pdf}
%%
%% Images with a different path to the parent latex file can
%% be accessed with the `import' package (which may need to be
%% installed) using
%%   \usepackage{import}
%% in the preamble, and then including the image with
%%   \import{<path to file>}{<filename>.pdf_tex}
%% Alternatively, one can specify
%%   \graphicspath{{<path to file>/}}
%% 
%% For more information, please see info/svg-inkscape on CTAN:
%%   http://tug.ctan.org/tex-archive/info/svg-inkscape
%%
\begingroup%
  \makeatletter%
  \providecommand\color[2][]{%
    \errmessage{(Inkscape) Color is used for the text in Inkscape, but the package 'color.sty' is not loaded}%
    \renewcommand\color[2][]{}%
  }%
  \providecommand\transparent[1]{%
    \errmessage{(Inkscape) Transparency is used (non-zero) for the text in Inkscape, but the package 'transparent.sty' is not loaded}%
    \renewcommand\transparent[1]{}%
  }%
  \providecommand\rotatebox[2]{#2}%
  \newcommand*\fsize{\dimexpr\f@size pt\relax}%
  \newcommand*\lineheight[1]{\fontsize{\fsize}{#1\fsize}\selectfont}%
  \ifx\svgwidth\undefined%
    \setlength{\unitlength}{218.54888435bp}%
    \ifx\svgscale\undefined%
      \relax%
    \else%
      \setlength{\unitlength}{\unitlength * \real{\svgscale}}%
    \fi%
  \else%
    \setlength{\unitlength}{\svgwidth}%
  \fi%
  \global\let\svgwidth\undefined%
  \global\let\svgscale\undefined%
  \makeatother%
  \begin{picture}(1,1.13324408)%
    \lineheight{1}%
    \setlength\tabcolsep{0pt}%
    \put(0,0){\includegraphics[width=\unitlength,page=1]{lab_setup_2_svg-tex.pdf}}%
    \put(0.27488614,1.02646289){\makebox(0,0)[t]{\lineheight{0.75}\smash{\begin{tabular}[t]{c}Thermal imaging\\camera\end{tabular}}}}%
    \put(0.14922691,0.3025247){\makebox(0,0)[t]{\lineheight{0.75}\smash{\begin{tabular}[t]{c}Beam, TEMs and\\electronics\end{tabular}}}}%
    \put(0.16385273,0.58287771){\makebox(0,0)[t]{\lineheight{0.75}\smash{\begin{tabular}[t]{c}Water cooling\\system\end{tabular}}}}%
    \put(0,0){\includegraphics[width=\unitlength,page=2]{lab_setup_2_svg-tex.pdf}}%
  \end{picture}%
\endgroup%

%% file: boundary_control_dist_highlight_hint_svg-tex.pdf_tex
%% Creator: Inkscape 1.3.2 (1:1.3.2+202311252150+091e20ef0f), www.inkscape.org
%% PDF/EPS/PS + LaTeX output extension by Johan Engelen, 2010
%% Accompanies image file 'boundary_control_dist_highlight_hint_svg-tex.pdf' (pdf, eps, ps)
%%
%% To include the image in your LaTeX document, write
%%   \input{<filename>.pdf_tex}
%%  instead of
%%   \includegraphics{<filename>.pdf}
%% To scale the image, write
%%   \def\svgwidth{<desired width>}
%%   \input{<filename>.pdf_tex}
%%  instead of
%%   \includegraphics[width=<desired width>]{<filename>.pdf}
%%
%% Images with a different path to the parent latex file can
%% be accessed with the `import' package (which may need to be
%% installed) using
%%   \usepackage{import}
%% in the preamble, and then including the image with
%%   \import{<path to file>}{<filename>.pdf_tex}
%% Alternatively, one can specify
%%   \graphicspath{{<path to file>/}}
%% 
%% For more information, please see info/svg-inkscape on CTAN:
%%   http://tug.ctan.org/tex-archive/info/svg-inkscape
%%
\begingroup%
  \makeatletter%
  \providecommand\color[2][]{%
    \errmessage{(Inkscape) Color is used for the text in Inkscape, but the package 'color.sty' is not loaded}%
    \renewcommand\color[2][]{}%
  }%
  \providecommand\transparent[1]{%
    \errmessage{(Inkscape) Transparency is used (non-zero) for the text in Inkscape, but the package 'transparent.sty' is not loaded}%
    \renewcommand\transparent[1]{}%
  }%
  \providecommand\rotatebox[2]{#2}%
  \newcommand*\fsize{\dimexpr\f@size pt\relax}%
  \newcommand*\lineheight[1]{\fontsize{\fsize}{#1\fsize}\selectfont}%
  \ifx\svgwidth\undefined%
    \setlength{\unitlength}{721.94631021bp}%
    \ifx\svgscale\undefined%
      \relax%
    \else%
      \setlength{\unitlength}{\unitlength * \real{\svgscale}}%
    \fi%
  \else%
    \setlength{\unitlength}{\svgwidth}%
  \fi%
  \global\let\svgwidth\undefined%
  \global\let\svgscale\undefined%
  \makeatother%
  \begin{picture}(1,0.47365791)%
    \lineheight{1}%
    \setlength\tabcolsep{0pt}%
    \put(0,0){\includegraphics[width=\unitlength,page=1]{boundary_control_dist_highlight_hint_svg-tex.pdf}}%
    \put(0.28965368,0.15590669){\color[rgb]{0,0,0}\makebox(0,0)[t]{\lineheight{1.25}\smash{\begin{tabular}[t]{c}$0$\end{tabular}}}}%
    \put(0.65443132,0.04935249){\color[rgb]{0,0,0}\makebox(0,0)[t]{\lineheight{1.25}\smash{\begin{tabular}[t]{c}$L$\end{tabular}}}}%
    \put(0.74699881,0.03210834){\color[rgb]{0,0,0}\makebox(0,0)[t]{\lineheight{1.25}\smash{\begin{tabular}[t]{c}$\zeta$\end{tabular}}}}%
    \put(0,0){\includegraphics[width=\unitlength,page=2]{boundary_control_dist_highlight_hint_svg-tex.pdf}}%
    \put(0.92340633,0.11853635){\color[rgb]{0,0,0}\makebox(0,0)[t]{\lineheight{1.25}\smash{\begin{tabular}[t]{c}$I_{\psi,1}(t)$\end{tabular}}}}%
    \put(0.07096683,0.3639721){\color[rgb]{0,0,0}\makebox(0,0)[t]{\lineheight{1.25}\smash{\begin{tabular}[t]{c}$I_{\psi,0}(t)$\end{tabular}}}}%
    \put(0,0){\includegraphics[width=\unitlength,page=3]{boundary_control_dist_highlight_hint_svg-tex.pdf}}%
    \put(0.38930276,0.39058085){\color[rgb]{0,0,0}\makebox(0,0)[t]{\lineheight{1.25}\smash{\begin{tabular}[t]{c}$I_0(t)$\end{tabular}}}}%
    \put(0.83908266,0.26024464){\color[rgb]{0,0,0}\makebox(0,0)[t]{\lineheight{1.25}\smash{\begin{tabular}[t]{c}$I_1(t)$\end{tabular}}}}%
    \put(0,0){\includegraphics[width=\unitlength,page=4]{boundary_control_dist_highlight_hint_svg-tex.pdf}}%
  \end{picture}%
\endgroup%

%% file: control_loop_svg-tex.pdf_tex
%% Creator: Inkscape 1.3.2 (1:1.3.2+202311252150+091e20ef0f), www.inkscape.org
%% PDF/EPS/PS + LaTeX output extension by Johan Engelen, 2010
%% Accompanies image file 'control_loop_svg-tex.pdf' (pdf, eps, ps)
%%
%% To include the image in your LaTeX document, write
%%   \input{<filename>.pdf_tex}
%%  instead of
%%   \includegraphics{<filename>.pdf}
%% To scale the image, write
%%   \def\svgwidth{<desired width>}
%%   \input{<filename>.pdf_tex}
%%  instead of
%%   \includegraphics[width=<desired width>]{<filename>.pdf}
%%
%% Images with a different path to the parent latex file can
%% be accessed with the `import' package (which may need to be
%% installed) using
%%   \usepackage{import}
%% in the preamble, and then including the image with
%%   \import{<path to file>}{<filename>.pdf_tex}
%% Alternatively, one can specify
%%   \graphicspath{{<path to file>/}}
%% 
%% For more information, please see info/svg-inkscape on CTAN:
%%   http://tug.ctan.org/tex-archive/info/svg-inkscape
%%
\begingroup%
  \makeatletter%
  \providecommand\color[2][]{%
    \errmessage{(Inkscape) Color is used for the text in Inkscape, but the package 'color.sty' is not loaded}%
    \renewcommand\color[2][]{}%
  }%
  \providecommand\transparent[1]{%
    \errmessage{(Inkscape) Transparency is used (non-zero) for the text in Inkscape, but the package 'transparent.sty' is not loaded}%
    \renewcommand\transparent[1]{}%
  }%
  \providecommand\rotatebox[2]{#2}%
  \newcommand*\fsize{\dimexpr\f@size pt\relax}%
  \newcommand*\lineheight[1]{\fontsize{\fsize}{#1\fsize}\selectfont}%
  \ifx\svgwidth\undefined%
    \setlength{\unitlength}{282.43942141bp}%
    \ifx\svgscale\undefined%
      \relax%
    \else%
      \setlength{\unitlength}{\unitlength * \real{\svgscale}}%
    \fi%
  \else%
    \setlength{\unitlength}{\svgwidth}%
  \fi%
  \global\let\svgwidth\undefined%
  \global\let\svgscale\undefined%
  \makeatother%
  \begin{picture}(1,0.57757478)%
    \lineheight{1}%
    \setlength\tabcolsep{0pt}%
    \put(0,0){\includegraphics[width=\unitlength,page=1]{control_loop_svg-tex.pdf}}%
    \put(0.34942387,0.43348817){\makebox(0,0)[t]{\lineheight{1.25}\smash{\begin{tabular}[t]{c}Linearization\end{tabular}}}}%
    \put(0.34942386,0.39512326){\makebox(0,0)[t]{\lineheight{1.25}\smash{\begin{tabular}[t]{c}\eqref{eq:compensation}\end{tabular}}}}%
    \put(0.42762339,0.15749003){\makebox(0,0)[t]{\lineheight{1.25}\smash{\begin{tabular}[t]{c}ASMC\end{tabular}}}}%
    \put(0.42762339,0.11912511){\makebox(0,0)[t]{\lineheight{1.25}\smash{\begin{tabular}[t]{c}\eqref{eq:control_error_full}\end{tabular}}}}%
    \put(0.68091437,0.43348817){\makebox(0,0)[t]{\lineheight{1.25}\smash{\begin{tabular}[t]{c}Plant\end{tabular}}}}%
    \put(0.68091437,0.39512326){\makebox(0,0)[t]{\lineheight{1.25}\smash{\begin{tabular}[t]{c}\eqref{eq:PDEmodel}, \eqref{eq:BCsmodel}\end{tabular}}}}%
    \put(0.12214352,0.00670538){\makebox(0,0)[t]{\lineheight{1.25}\smash{\begin{tabular}[t]{c}$u_{i, \text{e}}$\end{tabular}}}}%
    \put(0.90738449,0.00668011){\makebox(0,0)[t]{\lineheight{1.25}\smash{\begin{tabular}[t]{c}$\Te(\zeta_i)$\end{tabular}}}}%
    \put(0.93634397,0.09022811){\makebox(0,0)[t]{\lineheight{1.25}\smash{\begin{tabular}[t]{c}$-$\end{tabular}}}}%
    \put(0.24195207,0.16740335){\makebox(0,0)[t]{\lineheight{1.25}\smash{\begin{tabular}[t]{c}$v_i(t)$\end{tabular}}}}%
    \put(0.70069423,0.16957544){\makebox(0,0)[t]{\lineheight{1.25}\smash{\begin{tabular}[t]{c}$\bar{z}(\zeta_i, t)$\end{tabular}}}}%
    \put(0.85911453,0.44557358){\makebox(0,0)[t]{\lineheight{1.25}\smash{\begin{tabular}[t]{c}$T(\zeta_i, t)$\end{tabular}}}}%
    \put(0.68329794,0.53703116){\makebox(0,0)[t]{\lineheight{1.25}\smash{\begin{tabular}[t]{c}$\psi_i(t)$\end{tabular}}}}%
    \put(0,0){\includegraphics[width=\unitlength,page=2]{control_loop_svg-tex.pdf}}%
    \put(0.52105784,0.44665771){\makebox(0,0)[t]{\lineheight{1.25}\smash{\begin{tabular}[t]{c}$I_i(t)$\end{tabular}}}}%
    \put(0.15982143,0.44871237){\makebox(0,0)[t]{\lineheight{1.25}\smash{\begin{tabular}[t]{c}$u_i(t)$\end{tabular}}}}%
    \put(0,0){\includegraphics[width=\unitlength,page=3]{control_loop_svg-tex.pdf}}%
  \end{picture}%
\endgroup%

%% file: root.bbl
\begin{thebibliography}{10}

\bibitem{Boskovic2002}
D.~M. Bošković and M.~Krstić, ``Backstepping control of chemical tubular reactors,'' {\em Computers \& Chemical Engineering}, vol.~26, pp.~1077--1085, Aug. 2002.

\bibitem{Forman2011}
J.~C. Forman, S.~Bashash, J.~L. Stein, and H.~K. Fathy, ``Reduction of an electrochemistry-based li-ion battery model via quasi-linearization and padé approximation,'' {\em Journal of The Electrochemical Society}, vol.~158, no.~2, p.~A93, 2011.

\bibitem{Eleiwi2017}
F.~Eleiwi and T.~M. Laleg-Kirati, ``Observer-based perturbation extremum seeking control with input constraints for direct-contact membrane distillation process,'' {\em Int. J. Contr.}, vol.~91, pp.~1363--1375, May 2017.

\bibitem{Burns2016}
J.~A. Burns, X.~He, and W.~Hu, ``Feedback stabilization of a thermal fluid system with mixed boundary control,'' {\em Computers and Mathematics with Applications}, vol.~71, no.~11, pp.~2170--2191, 2016.

\bibitem{Ng2013}
J.~Ng and S.~Dubljevic, ``Boundary control synthesis for a lithium-ion battery thermal regulation problem,'' {\em AIChE Journal}, vol.~59, no.~10, pp.~3782--3796, 2013.

\bibitem{Pisano2017}
A.~Pisano and Y.~Orlov, ``On the {ISS} properties of a class of parabolic dps’ with discontinuous control using sampled-in-space sensing and actuation,'' {\em Automatica}, vol.~81, pp.~447--454, July 2017.

\bibitem{Wang2019}
J.-J. Gu and J.-M. Wang, ``Sliding mode control for n-coupled reaction-diffusion pdes with boundary input disturbances,'' {\em International Journal of Robust and Nonlinear Control}, vol.~29, no.~5, pp.~1437--1461, 2019.

\bibitem{Koch2022}
S.~Koch, A.~Pilloni, A.~Pisano, and E.~Usai, ``Sliding-mode boundary control of an in-line heating system governed by coupled {PDE}/{ODE} dynamics,'' {\em IEEE Trans. Contr. Syst. Tech.}, vol.~30, no.~6, pp.~2689--2697, 2022.

\bibitem{Zhou2025}
W.-J. Zhou, K.-N. Wu, and Y.-G. Niu, ``Robust sliding mode boundary stabilization for uncertain delay reaction–diffusion systems,'' {\em IEEE Trans. Aut. Contr.}, vol.~70, no.~1, p.~549–556, 2025.

\bibitem{Balogoun2025}
I.~Balogoun, S.~Marx, and F.~Plestan, ``Sliding mode control for a class of linear infinite-dimensional systems,'' {\em IEEE Trans. Aut. Contr.}, vol.~70, no.~5, p.~3464–3470, 2025.

\bibitem{Zhang2025}
J.~Zhang and W.~Wu, ``Robust sliding mode control for a class of gantry crane system with time‐varying disturbances,'' {\em International Journal of Robust and Nonlinear Control}, vol.~35, no.~8, p.~3055–3070, 2025.

\bibitem{Plestan2010}
F.~Plestan, Y.~Shtessel, V.~Brégeault, and A.~Poznyak, ``New methodologies for adaptive sliding mode control,'' {\em Int. J. Contr.}, vol.~83, no.~9, pp.~1907--1919, 2010.

\bibitem{Mayr2024}
P.~Mayr, Y.~Orlov, A.~Pisano, S.~Koch, and M.~Reichhartinger, ``Adaptive sliding mode boundary control of a perturbed diffusion process,'' {\em Int. J. Rob. Nonlin. Control}, vol.~34, no.~15, pp.~10055--10067, 2024.

\bibitem{Roy2020}
S.~Roy, S.~Baldi, and L.~M. Fridman, ``On adaptive sliding mode control without a priori bounded uncertainty,'' {\em Automatica}, vol.~111, p.~108650, Jan. 2020.

\bibitem{Han2024}
Z.~Han, Z.~Liu, J.-W. Wang, and W.~He, ``Fault-tolerant control for flexible structures with partial output constraint,'' {\em IEEE Trans. Aut. Contr.}, vol.~69, no.~4, pp.~2668--2675, 2024.

\bibitem{Lineykin2007}
S.~Lineykin and S.~Ben-Yaakov, ``Modeling and analysis of thermoelectric modules,'' {\em IEEE Transactions on Industry Applications}, vol.~43, no.~2, pp.~505--512, 2007.

\bibitem{Orlov2022book}
Y.~Orlov, {\em Nonsmooth Lyapunov Analysis in Finite and Infinite Dimensions}.
\newblock Springer, 2022.

\bibitem{Krasnoselskii1976book}
M.~Krasnoselskii, {\em Integral Operators in Spaces of Summable Functions}.
\newblock Noordhoff, 1976.

\bibitem{Smyshlyaev2010}
A.~Smyshlyaev and M.~Krstic, {\em Adaptive Control of Parabolic PDEs.}
\newblock Princeton: Princeton University Press, 2010.

\bibitem{Khalil2002}
H.~Khalil, {\em Nonlinear Systems}.
\newblock Prentice Hall, 2002.

\bibitem{Brezis2010}
H.~Brezis, {\em Functional Analysis, Sobolev Spaces and Partial Differential Equations}.
\newblock Springer New York, 2010.

\bibitem{Zaitsev2002}
V.~F. Zaitsev, {\em Handbook of Exact Solutions for Ordinary Differential Equations}.
\newblock London: CRC Press, 2nd~ed., 2002.

\end{thebibliography}
